\documentclass[%
 preprint, 
 amsmath,amssymb,
 aps, physrev,
]{revtex4-2}

\usepackage{graphicx}
\usepackage{dcolumn}
\usepackage{bm}
\usepackage{mathrsfs}
\usepackage{amsmath}
\usepackage{xcolor}
\usepackage{subfigure}
\usepackage{graphicx}
\usepackage{tabularray}
\usepackage{hyperref}
\usepackage{enumitem}
\usepackage{placeins}

\begin{document}

\preprint{ }

\title{\textbf{Quasinormal modes of the thick braneworld in $f(T)$ gravity} 
}%

\author{Zi-Jie Li$^{a}$$^{b}$}
\author{Hai-Long Jia$^{a}$$^{b}$}
\author{Qin Tan$^{c}$}
\author{Wen-Di Guo$^{a}$$^{b}$}
	\altaffiliation{guowd@lzu.edu.cn, corresponding author}

\affiliation{
	$^{a}$Lanzhou Center for Theoretical Physics, 
	Key Laboratory of Theoretical Physics of Gansu Province,
	Key Laboratory of Quantum Theory and Applications of MoE,
	Gansu Provincial Research Center for Basic Disciplines of Quantum Physics, Lanzhou 
	University, Lanzhou 730000, China \\
	$^{b}$Institute of Theoretical Physics \& Research Center of Gravitation,
	School of Physical Science and Technology, Lanzhou University, Lanzhou 730000, 
	China \\
	$^{c}$Department of Physics,
	Key Laboratory of Low Dimensional Quantum Structures and Quantum Control of Ministry of Education,
	Synergetic Innovation Center for Quantum Effects and Applications,
	Hunan Normal University,
	Changsha, 410081, Hunan, China
}

\begin{abstract}
We investigate the quasinormal modes (QNMs) of a thick brane model in $f(T)$ gravity with $f(T) = T + \alpha T^2$. Requiring the energy density to remain positive and the scalar field to be real constrains the parameter $\alpha$ to the range $[-\frac{7}{48},\frac{1}{48}]$. Within this allowed region, we find that the parameter $\alpha$ can induce a brane-splitting structure. 
The quasinormal frequencies of the system are computed using both the asymptotic iteration method and the Bernstein spectral method. The two approaches show good agreement in the low-overtone regime. For $\alpha<0$, the decay rate of the first QNM decreases as $|\alpha|$ increases, whereas higher overtones exhibit the opposite behavior. 
To further examine the influence of model parameters on the QNM spectrum, we also perform numerical time-domain evolution of perturbations, whose results are consistent with the frequency-domain analysis. Our results provide a concrete example of quasinormal spectra in thick brane models within $f(T)$ gravity and may offer useful insights for future observational tests of extra dimensions. 

\end{abstract}

\maketitle{}

\section{Introduction}
The idea of extra dimensions was first proposed by Gunnar Nordstr\"{o}m \cite{Nordstrom}. Later, inspired by the general relativity and guided by the experimental constraints, the Kaluza-Klein (KK) theory was developed, in which the extra dimension is assumed to be compactified on a circle \cite{Kaluza,Klein}. In this framework, particles can possess nonvanishing momentum components along the fifth dimension, giving rise to the so-called KK modes. These modes propagate in the extra dimension, while their four-dimensional projections are interpreted as particles observed in our universe.\par

With the subsequent discovery of the strong and weak interactions, the idea of higher-dimensional physics was further extended. An early braneworld scenario was proposed by Akama \cite{Akama}, in which our observable universe is regarded as a hypersurface embedded in a higher-dimensional spacetime. Later, Arkani-Hamed, Dimopoulos, and Dvali proposed the ADD model to address the hierarchy problem between the gravitational scale and electroweak scale \cite{ADD}. However, this model effectively shifts the hierarchy problem to another hierarchy between the size of the extra dimension and the fundamental gravitational scale in higher dimensions. 
In 1999, Randall and Sundrum proposed the RS-\uppercase\expandafter{\romannumeral1} model, which provides a novel mechanism to resolve the hierarchy problem through a warped extra dimension \cite{Randall1}. Soon after, the RS-\uppercase\expandafter{\romannumeral2} model was introduced, demonstrating that four-dimensional Newtonian gravity can be recovered on the brane even when the extra dimension is infinite \cite{Randall2}. Since then,  braneworld models have been widely applied in various areas of theoretical physics \cite{Shiromizu,Tanaka,Gregory,Jaman,Adhikari,Geng:2020fxl,Geng:2021iyq,Geng:2022dua,Bhattacharya:2021jrn}. 
However, the RS-\uppercase\expandafter{\romannumeral2} model treats the brane as an infinitely thin hypersurface and neglects its internal structure. To construct more realistic scenarios, thick brane models were developed by introducing scalar, vector, or spinor fields to generate a finite brane thickness \cite{Rubakov:1983kt,DeWolfe:1999cp,Gremm:1999pj,Csaki:2000fc,Dzhunushaliev,Dzhunushaliev:2011mm}. Thick branes arise naturally in many gravitational theories and have been extensively investigated in different modified gravity frameworks \cite{Afonso:2007gc,Liu:2011wi,Bazeia:2013uva,Geng,Gu:2016nyo,Zhong:2016iko,Zhong:2017ffr,Zhou:2017xaq,Xie:2021ayr,Moreira:2021uod,Xu:2022xxd,Silva:2022pfd,Xu:2022gth}. The finite thickness of the brane can lead to a variety of interesting physical properties and phenomenological. Comprehensive reviews on braneworld models can be found in Refs. \cite{Dzhunushaliev:2009va,Maartens:2010ar,Liu:2017gcn,Ahluwalia:2022ttu}.

In conservative systems, the characteristic oscillation modes are known as normal modes, whereas in dissipative systems they are referred to as quasinormal modes (QNMs). QNMs provide an effective tool for probing the properties of physical systems and have been widely studied in black hole physics \cite{Kokkotas:1999bd,Nollert:1999ji,Berti:2009kk,Konoplya:2011qq,Cardoso:2016rao,Jusufi:2020odz,Cheung:2021bol}. Since thick brane models also represent dissipative systems, they naturally possess quasinormal spectra \cite{Seahra:2005iq,Seahra:2005wk,Deng1,Yang1,Jia1,Jia2,Guo1,Guo2,Tan:2022uex,Tan5,Tan4,Tan3,Tan2,Tan1,Zhu:2024gvl,E:2025kic}. By analyzing the QNMs, one can extract characteristic signatures of the thick brane, in much the same way as the parameters of a black hole can be inferred from its QNMs. Therefore, investigating QNMs in braneworld scenarios is of considerable interest.

General relativity describes gravity in terms of spacetime curvature. In contrast, the teleparallel equivalent of general relativity (TEGR) formulates gravity using spacetime torsion \cite{Hayashi:1979wj}. Although the two are equivalent, their extensions can lead to different physical consequences. In particular, when TEGR is generalized to $f(T)$ gravity, torsional effects become dynamical and can significantly modify the gravitational theory \cite{Aldrovandi:2013wha}. Such models have attracted considerable attention, especially in attempts to explain the dark energy problem \cite{Ferraro:2006jd,Bengochea:2008gz,Linder:2010py,Karami:2010bys,Bamba:2010wb}. For comprehensive discussions of $f(T)$ gravity, see the reviews in Refs. \cite{Cai,Krssak:2018ywd,Huguet:2020ler,Bahamonde:2021gfp}. 
Despite extensive studies of thick brane models in various modified gravity theories, investigations of thick branes in $f(T)$ gravity remain relatively limited \cite{Tan:2020sys}. Motivated by this, in this work we study the QNMs of a thick brane model within the framework of $f(T)$ gravity.

The remainder of this paper is organized as follows. In Sec. \ref{Secf(T)}, we briefly review the braneworld setup in $f(T)$ gravity. In Sec. \ref{SecQNMs}, we compute the quasinormal frequencies of the model using several methods and study the time evolution of incident wave packets through numerical simulations. Finally, Sec. \ref{SecCon} is devoted to discussions and conclusions.

\section{The braneworld in $f(T)$ gravity}
\label{Secf(T)}
In this section, we briefly review the thick braneworld solution in $f(T)$ gravity and its tensor perturbation. In teleparallel gravity,  it is convenient to work in the tangent space at each point of the spacetime manifold. For a point with spacetime coordinates $x^M$, one introduces a vielbein field $e_A(x^M)$, which forms an orthonormal basis in the tangent space. In this paper, we use capital Latin indices $A,B,C,D,\dots = 0,1,2,3,5$ to denote tangent-space coordinates, and capital Latin indices $M,N,O,P,\dots = 0,1,2,3,5$ to denote spacetime coordinates. For the four-dimensional brane, Greek indices $\mu,\nu,\sigma,\dots = 0,1,2,3$ denote spacetime coordinates, while lowercase Latin indices $a,b,c,d,\dots = 0,1,2,3$ denote the tangent-space coordinates. Lowercase Latin indices $i,j,\dots = 1,2,3$ label the three-dimensional space coordinates on the brane. The dual vector of $e_A(x^M)$ is written as $e^A(x^M)$, whose components are denoted by $e_A{}^M$ and $e^A{}_M$, respectively. Using the vielbein, the spacetime metric can be constructs as \begin{equation}\label{relation}
	g_{MN} = e_M{}^A e_N{}^B \eta_{AB},
\end{equation}
where $\eta_{AB}$ is the Minkowski metric in the tangent space. In teleparallel gravity, the gravitational interaction is described by the Weitzenb\"{o}ck connection, defined as
\begin{equation}
	\tilde{\Gamma}^O{}_{MN} = e_A{}^O \partial_N e^A{}_M.
\end{equation}
The torsion tensor is defined as
\begin{equation}
	T^O{}_{MN} = \tilde{\Gamma}^O{}_{NM} - \tilde{\Gamma}^O{}_{MN}.
\label{torsion}
\end{equation}
From the difference between the Weitzenb\"ock connection and the Levi--Civita connection $\Gamma^O{}_{ NM}$, one can define the contorsion tensor
\begin{equation}
	K^O{}_{MN} = \tilde{\Gamma}^O{}_{MN} - \Gamma^O{}_{NM}.
\label{contorsion}
\end{equation}
The superpotential tensor is then introduced as
\begin{equation}
	S_O{}^{MN} = \frac{1}{2} (K^{MN}{}_O - \delta^N_O T^{QM}{}_Q + \delta^M_O T^{QN}{}_Q),
\label{superpotential}
\end{equation}
where $\delta^N_O$ is the Kronecker $\delta$.
With these quantities, the torsion scalar can be constructed as
\begin{equation}
	T = S_O{}^{MN} T^O{}_{MN}.
\label{torsionscalar}
\end{equation}
The Lagrangian of teleparallel gravity is given by
\begin{equation}
	\mathscr{L}_T = -\frac{M^3_*}{4} e T,
\label{TEGRlagrangian}
\end{equation}
where $e = \text{det}(e^A_M) = \sqrt{-g}$ with $g$ being the determinant of the metric $g_{MN}$, and $M_*$ denotes the five-dimensional gravitational scale. The $f(T)$ gravity is a generalization of teleparallel gravity in which the torsion scalar $T$ in the Lagrangian is replaced by a general function $f(T)$, analogous to the extension from general relativity to $f(R)$ gravity. The corresponding Lagrangian density reads
\begin{equation}
	\mathscr{L}_{f(T)} = -\frac{M^3_*}{4} e f(T).
\label{fTlagrangian}
\end{equation}

\par
In this paper, we consider a braneworld model in five-dimensional $f(T)$ gravity. The action of the system is given by
\begin{equation}
	S = -\frac{M^3_*}{4} \int d^5 x e f(T) + \int d^5 x \mathscr{L}_m,
\label{action}
\end{equation}
where 
\begin{equation}
	\mathscr{L}_m = e \left( -\frac{1}{2} (\partial^M\phi)\partial_M\phi - V(\phi) \right)
\label{Lm}
\end{equation}
is the Lagrangian of a massless canonical scalar field which is commonly introduced to generate the thick brane configuration. Varying the action with respect to the vielbein and the scalar field yields the field equations
\begin{equation}
	e^{-1}f_Tg_{NP}\partial_Q(eS_M{}^{PQ})+f_{TT}S_{MN}{}^Q\partial_QT-f_T\tilde{\Gamma}^P{}_{QM}S_{PN}{}^Q+\frac{1}{4}g_{MN}f(T)=\mathscr{T}_{MN},
\label{FieldEquation}
\end{equation}
\begin{equation}
	\frac{1}{e} \partial^{M} \left( e\partial_{M}\phi \right) = \frac{dV}{d\phi},
\label{ScalarFieldEquation}
\end{equation}
where $f_T=\frac{df(T)}{dT}$, $f_{TT}=\frac{d^2f(T)}{dT^2}$, and $\mathscr{T}_{MN} = \left(\partial_M\phi \right) \partial_N\phi - g_{MN} \left( \frac{1}{2} g^{ST} (\partial_S\phi) \partial_T \phi + V(\phi)\right)$ denotes the energy-momentum tensor of the matter field. In the following, we set the five-dimensional gravitational scale $M_*=1$. To construct a thick brane solution, we adopt the metric ansatz \cite{Melfo:2002}
\begin{equation}
	ds^2=e^{2A(y)}\eta_{\mu\nu}dx^{\mu}dx^{\nu}+e^{2H(y)}dy^2,
\label{metricansatz}
\end{equation}
where $A(y)=\delta H(y)$. The corresponding vielbein can then be written as
\begin{equation}
	e^A{}_M=\text{diag}\left(e^A,e^A,e^A,e^A,e^H\right).
\label{vielbein}
\end{equation}
Substituting Eqs. \eqref{metricansatz} and \eqref{vielbein} into the field equations \eqref{FieldEquation} and \eqref{ScalarFieldEquation}, we obtain
\begin{equation}
	\frac{f}{4} + 6 A'^2 f_{T} e^{-2H} = -V + \frac{1}{2} \phi'^2 e^{-2H}, 
\label{55}
\end{equation}
\begin{equation}
	\frac{f}{4} + \frac{3}{2} \left(4A'^2 - A'H' + A''\right) f_{T} e^{-2H} + 36 A'^2 \left(A'H'-A''\right) f_{TT} e^{-4H} = -V -\frac{1}{2} \phi'^2 e^{-2H},
\label{11}
\end{equation}
\begin{equation}
	\phi'' + \left(4A' - H'\right)\phi' = e^{2H} \frac{dV}{d\phi},
\label{phi}
\end{equation}
where the prime ($'$) denotes differentiation with respect to $y$. From Eq. \eqref{55} and \eqref{11}, we obtain that
\begin{equation}
	\phi'^2 = \frac{3}{2} (A'H' - A'') f_T - 36A'^2 \left(A'H' - A''\right) f_{TT} e^{-2H},
\label{phieq}
\end{equation}
\begin{equation}
	V = -\frac{f}{4} - \frac{3}{4} (8A'^2 - A'H' + A'') f_T e^{-2H} - 18 A'^2 \left( A'H'-A'' \right) f_{TT} e^{-4H}.
\label{Veq}
\end{equation}
Only two of Eqs. \eqref{phi}, \eqref{phieq}, and \eqref{Veq} are independent, and we see that there are four unknown functions, $A(y)$, $f(T)$, $\phi(y)$, and $V(\phi)$. Therefore, two of the functions must be specified in order to determine the system completely. In this work, we choose the warp factor \cite{Melfo:2002}

\begin{equation}
	A(y) = -\frac{\delta}{2 s} \ln \left( 1 + \frac{ky}{\delta} \right)^{2s},
\end{equation}
and adopt the $f(T)$ model

\begin{equation}
	f(T) = T + \alpha T^2.
\end{equation}

\begin{figure}[htbp]
	\centering
	\subfigure[$s=1,\alpha=-1/200$]{\label{FigADelta}
		\includegraphics[width=0.3\textwidth]{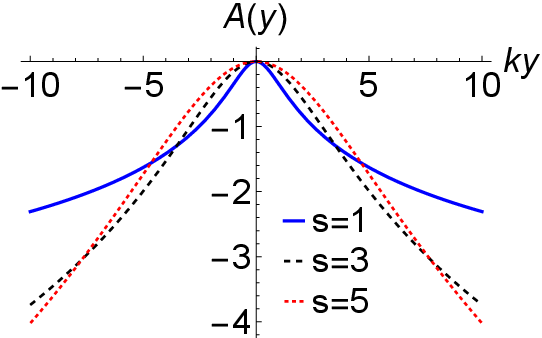}}
	\subfigure[$\delta=1,\alpha=-1/200$]{\label{FigAS}
		\includegraphics[width=0.3\textwidth]{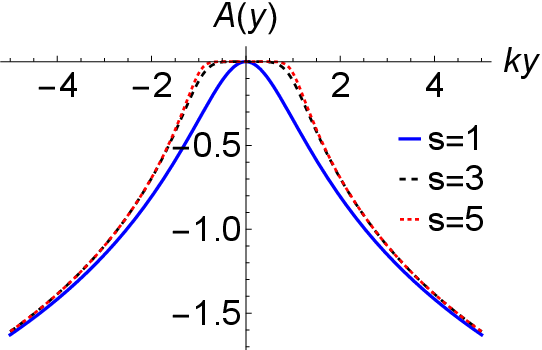}}
	\caption{Plots of the warp factor $A(y)$ for different parameters.}
	\label{FigA}
\end{figure}

Next, we briefly analyze the background configuration of the model, including the energy density, scalar field, and scalar  potential. The plots of $A(y)$ for different values of $\delta$ and $s$ are shown in Fig. \ref{FigA}. From Fig. \ref{FigAS}, one can observe that a platform structure appears near $y=0$ when $s>1$. The energy density measured by a static observer is given by

\begin{figure}[htbp]
	\centering
	\subfigure[$s=1,\alpha=-1/200$]{\label{FigRhoDelta}
		\includegraphics[width=0.3\textwidth]{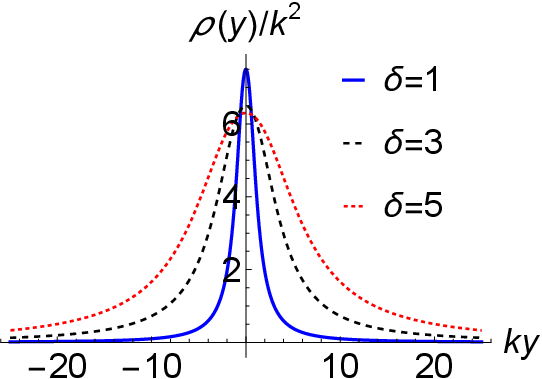}}
	\subfigure[$\delta=1,\alpha=-1/200$]{\label{FigRhoS}
		\includegraphics[width=0.3\textwidth]{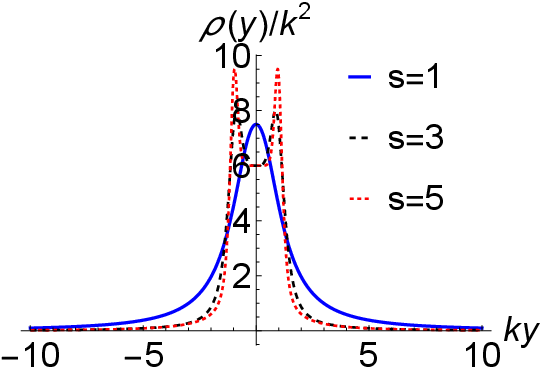}}
	\\
	\subfigure[$s=1,\delta=1$]{\label{FigRhoAlpha}
		\includegraphics[width=0.3\textwidth]{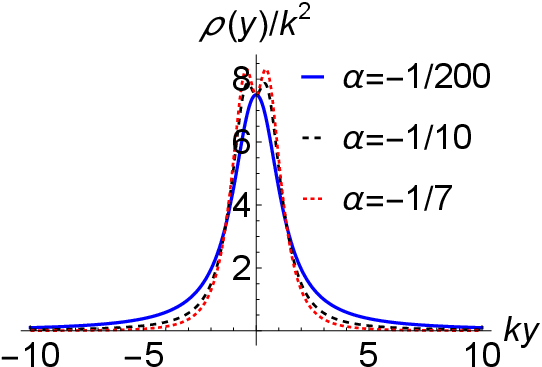}}
	\subfigure[$s=1,\delta=1$]{\label{FigRhoAlphaL0}
		\includegraphics[width=0.3\textwidth]{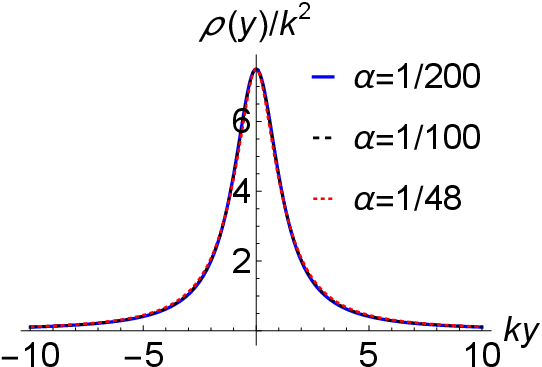}}
	\caption{Plots of the energy density for different parameters.}
	\label{FigRho}
\end{figure}

\begin{equation}\label{EnergyDensity}
	\begin{split}
		\rho(y) &= T_{MN}Z^MZ^N\\
		&= -\frac{1}{4}f - \frac{3}{2} f_T e^{-2 H(y)} \left(A''(y)-A'(y) H'(y)+4 A'(y)^2\right)\\
		&+36 f_{TT} e^{-4 H(y)} A'(y)^2
		\left(A''(y)-A'(y) H'(y)\right),
	\end{split}
\end{equation}
where $Z^M = (e^{-A},0,0,0,0)$ is the five-dimensional velocity of a static observer. Notice that Eq. \eqref{EnergyDensity}  approaches a negative constant at large $|y|$. This behavior arises because the spacetime is asymptotically anti-de Sitter, which introduces a negative cosmological constant. In order to analyze the physical brane structure, the contribution from this background cosmological constant should be subtracted. 
The energy density under different parameters are shown in Fig. \ref{FigRho}. From \ref{FigRhoDelta}, we observe that when $\delta$ increases, the peak of $\rho$ becomes lower and broader, indicating that the thickness of the brane increases. For $s>1$, the brane splits into two sub-branes, as illustrated in Fig. \ref{FigRhoS}. The similar splitting phenomenon can also occur when $\alpha$ is sufficiently small, as shown in Fig. \ref{FigRhoAlpha}. However, they are different. The splitting caused by $s$ produces a platform structure around $y=0$, whereas the splitting induced by $\alpha$ does not exhibit such a platform. 
It is also worth noting that for certain values of $\alpha$, the energy density becomes negative and the scalar field develops an imaginary component, which is physically unacceptable. To ensure that the energy density remains positive and that the scalar field is real, the parameter $\alpha$ must lie within the range $[-\frac{7}{48},\frac{1}{48}]$. When $\alpha > 0$, the corresponding energy density plots are shown in Fig. \ref{FigRhoAlphaL0}, where no particularly interesting structure appear. Therefore, in the following analysis, we mainly focus on the case where $\alpha<0$.

\begin{figure}[htbp]
	\centering
	\subfigure[$s=1,\alpha=-1/200$]{\label{FigPhiDelta}
		\includegraphics[width=0.3\textwidth]{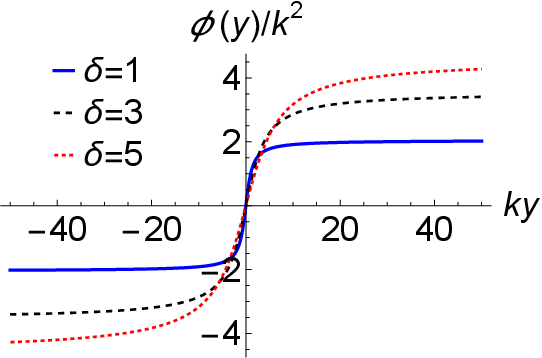}}
	\subfigure[$\delta=1,\alpha=-1/200$]{\label{FigPhiS}
		\includegraphics[width=0.3\textwidth]{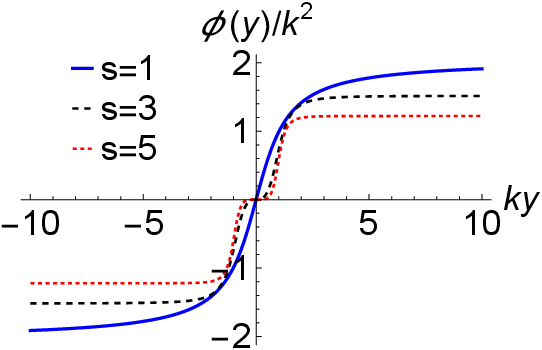}}
	\subfigure[$s=1,\delta=1$]{\label{FigPhiAlpha}
		\includegraphics[width=0.3\textwidth]{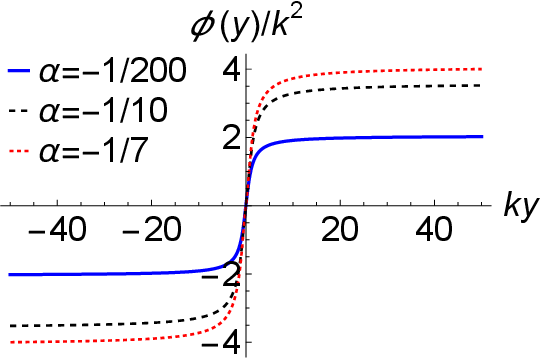}}
	\caption{Plots of the scalar field for different parameters.}
	\label{FigPhi}
\end{figure}

\begin{figure}[htbp]
	\centering
	\subfigure[$s=1,\alpha=-1/200$]{\label{FigVDelta}
		\includegraphics[width=0.3\textwidth]{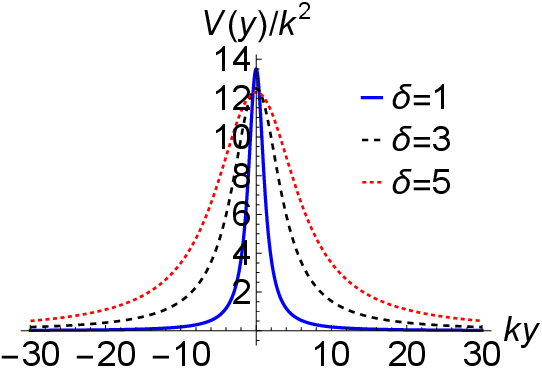}}
	\subfigure[$\delta=1,\alpha=-1/200$]{\label{FigVS}
		\includegraphics[width=0.3\textwidth]{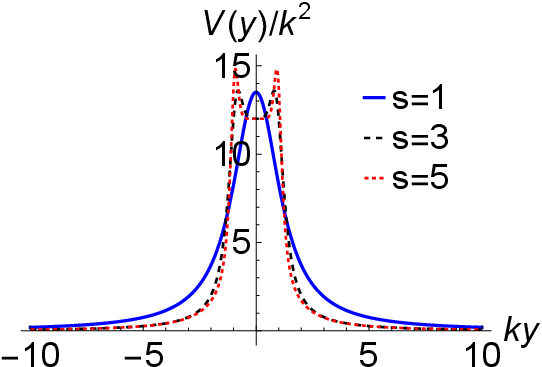}}
	\subfigure[$s=1,\delta=1$]{\label{FigVAlpha}
		\includegraphics[width=0.3\textwidth]{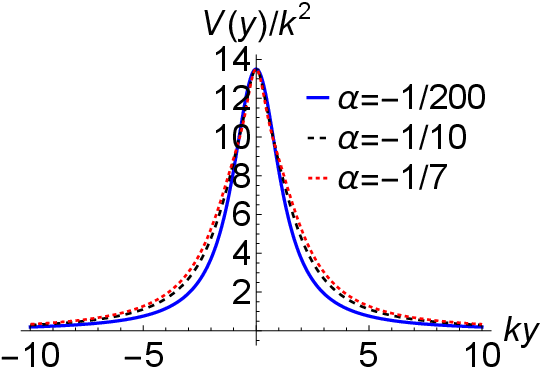}}
	\caption{Plots of the scalar potential for different parameters.}
	\label{FigV}
\end{figure}

The plots of scalar field and scalar potential for different parameters are as shown in Figs. \ref{FigPhi} and \ref{FigV}. From Fig. \ref{FigPhiS}, we observe that when $s>1$, the scalar field develops a kink-like structure around $y = 0$. However, this feature does not appear in Fig. \ref{FigPhiAlpha}. This kink structure of $\phi$ is closely related to the platform structure observed in the plot of energy density $\rho$. In addition, the asymptotic value of $\phi$ as $y \rightarrow \infty$ increases when $s$ or $\alpha$ decreases, or when $\delta$ increases. The splitting phenomenon is observed again from Fig. \ref{FigVS} while does not occur in Fig. \ref{FigVAlpha}.  

\par
By performing the coordinate transformation $d\bar{y}=e^Hdy$, the metric can be rewritten as
\begin{equation}\label{Metric}
	ds^2=e^{2A(\bar{y})}\eta_{\mu\nu}dx^{\mu}dx^{\nu}+d\bar{y}^2
\end{equation}
with the corresponding vielbein
\begin{equation}\label{Vielbein}
	e^A{}_M=\text{diag}\left(e^A,e^A,e^A,e^A,1\right).
\end{equation}

Tensor perturbation can decouple from vector and scalar perturbations, so we can study them separately. In this paper, we only study tensor perturbation. The perturbed vielbein is written as
\begin{equation}
	e^A{}_M =
	\begin{pmatrix}
		e^{A(\bar{y})}(\delta^a{}_\mu + h^a{}_\mu)  & 0\\
		0                                      & 1 
	\end{pmatrix},
\end{equation}
where $h^a{}_\mu$ represents the tensor perturbation which is a function defined on five-dimensional spacetime coordinates. Using the relation between the metric and the vielbein given by Eq. \eqref{relation}, the perturbed metric is
\begin{equation}
	g_{MN} = 
	\begin{pmatrix}
		e^{2A(\bar{y})}\left(\eta_{\mu\nu} + \gamma_{\mu\nu}\right) & 0\\
		0                                          & 1
	\end{pmatrix},
\end{equation}
where 
\begin{equation}
	\gamma_{\mu\nu} = (\delta^a{}_\mu h^b{}_\nu + \delta^b{}_\nu h^a{}_\mu) \eta_{ab}.
\end{equation}
Keeping only the first-order infinitesimal quantities, the inverse veilbein and inverse metric are given by
\begin{equation}
	e_A{}^M = 
	\begin{pmatrix}
		e^{-A(\bar{y})}\left(\delta_a{}^\mu-h_a{}^\mu\right) & 0\\
		0                                   & 1    
	\end{pmatrix}
\end{equation}
and
\begin{equation}
	g^{MN} = 
	\begin{pmatrix}
		e^{-2A(\bar{y})}\left(\eta^{\mu\nu} - \gamma^{\mu\nu}\right) & 0\\
		0
		& 1
	\end{pmatrix}
\end{equation}
respectively, where
\begin{equation}
	\gamma^{\mu\nu} = (\delta_a{}^\mu h_b{}^\nu+\delta_b{}^\nu h_a{}^\mu) \eta^{ab}.
\end{equation}
The tensor perturbation $\gamma_{\mu\nu}$ satisfies the transverse traceless condition:
\begin{equation}
	\partial_\mu \gamma^{\mu\nu} = 0,
	\qquad
	\eta^{\mu\nu} \gamma_{\mu\nu} = 0,
\end{equation}
which are equivalent to
\begin{equation}
	\partial_\mu \left( \delta_a{}^\mu h_b{}^\nu + \delta_b{}^\nu h_a{}^\mu \right) \eta^{ab} = 0,
	\qquad
	\delta_a{}^\mu h^a{}_\mu = 0.
\end{equation}
After linearizing the field equation, the tensor perturbation satisfies
\begin{equation}
	\left( e^{-2A} \Box^{(4)} \gamma_{\mu\nu} + \partial^2_{\bar{y}}\gamma_{\mu\nu} + 4 \left(\partial_{\bar{y}}A\right) \partial_{\bar{y}}\gamma_{\mu\nu} \right)f_T - 24 \left(\partial_{\bar{y}}A\right) \left(\partial^2_{\bar{y}}A\right) \left(\partial_{\bar{y}}\gamma_{\mu\nu}\right) f_{TT} = 0,
\end{equation}
where $\Box^{4} \equiv \eta^{\mu\nu}\partial_{\mu}\partial_{\nu}  $.\par

For more details of the derivation, see Ref. \cite{guo2015}.
Introducing the conformal coordinate
\begin{equation}
	dz = e^{-A} d\bar{y},
\end{equation}
the perturbation equation can be rewritten as
\begin{equation}
	\left( \partial^2_z + 2H\partial_z + \Box^{(4)} \right)\gamma_{\mu\nu}=0,
\end{equation}
where 
\begin{equation}
	H = \frac{3}{2} \partial_z A + 12 e^{-2A}\left( (\partial_z A)^3 - (\partial^2_zA)\partial_zA \right)\frac{f_{TT}}{f_T}.
\end{equation}\par

To obtain the dynamical equation of perturbation along the extra dimension, we perform the KK decomposition, which can be regarded as an application of the method of separation of variables \cite{Seahra:2005iq,guo2015},
\begin{equation}
	\gamma_{\mu\nu}(x^{\rho}, z) = e^{-\frac{3}{2}A(z)} e^{\int K(z) dz} \psi(z, t) e^{-i a_i x^i} \epsilon_{\mu\nu},
\end{equation}
where $a_i x^i = \delta_{ij}a^i x^j$, $\epsilon_{\mu\nu} = \text{const.}$, and 
\begin{equation}
	K(z) = 12 e^{-2A} \left( (\partial^2_z A)\partial_z A - (\partial_z A)^3 \right) \frac{f_{TT}}{f_T}.
\end{equation}
Substituting the above decomposition into the perturbation equation, we obtain
\begin{equation}\label{eq_tz}
	-\partial^2_t \psi + \left( \partial^2_z - U_{\text{eff}}(z) \right)\psi - a^2 \psi = 0,
\end{equation}
where
$a^2 = \delta_{ij} a^i a^j$ and
\begin{equation}\label{Ueff}
	U_{\text{eff}}(z) = H^2 + \partial_z H
\end{equation}
is the effective potential.
Introducing the separation
\begin{equation}\label{PsiToVarphi}
	\psi(z, t) = e^{-i\omega t} \varphi(z),
\end{equation}
we obtain the Schr\"odinger-like equation
\begin{equation}\label{eq_zeff}
	\left( -\partial^2_z + U_{\text{eff}}(z) \right)\varphi(z) = m^2 \varphi(z),
\end{equation}
where $m^2 = \omega^2 - a^2$. Here $m$, $\omega$, and $a$ represent the mass, angular frequency, and the magnitude of the three-dimensional momentum of the KK modes, respectively. Equation \eqref{eq_zeff} can be further written in a super-symmetric form,

\begin{equation}
	Q Q^\dagger \varphi(z) = m^2 \varphi(z),
\end{equation}
where 
\begin{equation}
	Q = \partial_z + H,
	\qquad
	Q^\dagger = -\partial_z + H.
\end{equation}
Correspondingly, one can construct another equation with a dual potential,

\begin{equation}\label{eq_zdual}
	Q^\dagger Q \hat{\varphi}(z) = \left( -\partial^2_z + U_{\text{dual}}(z) \right) \hat{\varphi}(z) = m^2 \hat\varphi(z),  
\end{equation}
where
\begin{equation}\label{Udual}
	U_{\text{dual}}(z) = H^2 - \partial_z H.
\end{equation}
The effective potential and the dual potential have the same spectrum for the massive QNMs according to the super-symmetric quantum mechanics \cite{Cooper:1994eh,Ge:2018vjq}, which provides us with a significant advantage to compute the QNFs.

\section{Quasinormal modes of the brane}
\label{SecQNMs}
In this section, we employ several methods to compute the quasinormal frequencies (QNFs) of the brane, which amounts to solving the eigenvalue equation \eqref{eq_zdual}. The methods used in this work include the asymptotic iteration method (AIM) \cite{AIM_2005}, the Bernstein spectral method (BSM) \cite{BSM_2023}, the direct integration method (DIM) \cite{DIM_2013}, and a method based on numerical evolution. These different approaches are adopted to provide cross-checks of the results.

To determine the QNFs of the brane, appropriate boundary conditions must be imposed. They are given by 

\begin{equation}\label{boundary_z}
	\hat{\varphi}(z)\propto
	\begin{cases}
		e^{imz}, & z\to\infty,\\
		e^{-imz}, & z\to-\infty.
	\end{cases}
\end{equation}
According to Eq. \eqref{PsiToVarphi}, these boundary conditions \eqref{boundary_z} correspond to purely outgoing waves at both spatial infinities. It is worth noting that an analytic relation between the coordinates $y$ and $z$ exists only when $\delta = 1$, in which case $y=z$. Therefore, in this work we mainly focus our analysis to the case $\delta = 1$.

\begin{figure}[htbp]
	\centering
	\subfigure[$s=1,\alpha=-1/200$]{\label{FigUeffDelta}
		\includegraphics[width=0.3\textwidth]{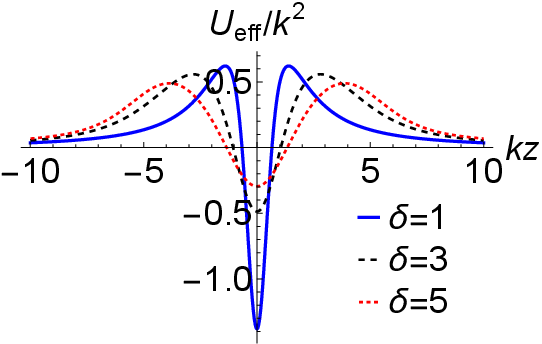}}
	\subfigure[$\delta=1,\alpha=-1/200$]{\label{FigUeffS}
		\includegraphics[width=0.3\textwidth]{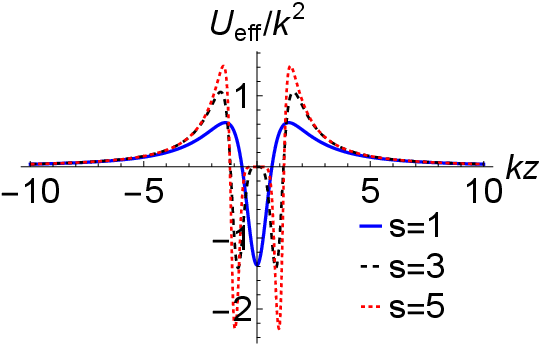}}
	\subfigure[$\delta=1,s=1$]{\label{FigUeffAlpha}
		\includegraphics[width=0.3\textwidth]{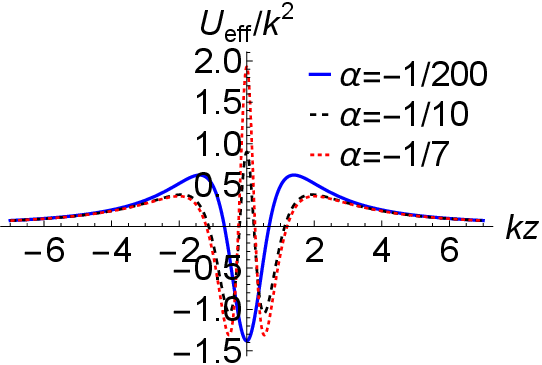}}
	\caption{Plots of the effective potential for different parameters.}
	\label{FigUeff}
\end{figure}

\begin{figure}[htbp]
	\centering
	\subfigure[$s=1,\alpha=-1/200$]{\label{FigUdualDelta}
		\includegraphics[width=0.3\textwidth]{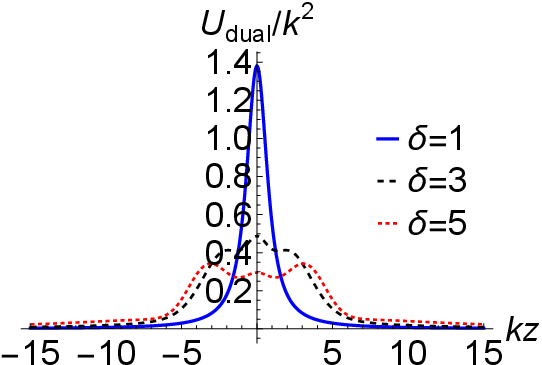}}
	\subfigure[$\delta=1,\alpha=-1/200$]{\label{FigUdualS}
		\includegraphics[width=0.3\textwidth]{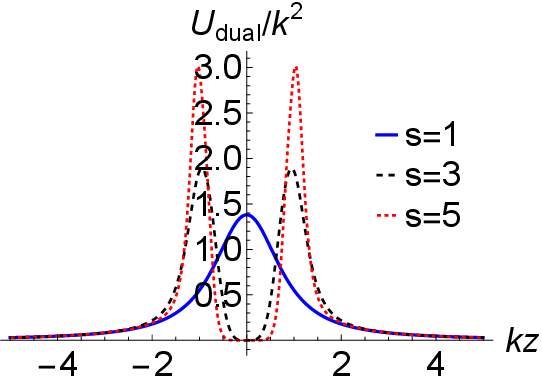}}
	\subfigure[$\delta=1,s=1$]{\label{FigUdualAlpha}
		\includegraphics[width=0.3\textwidth]{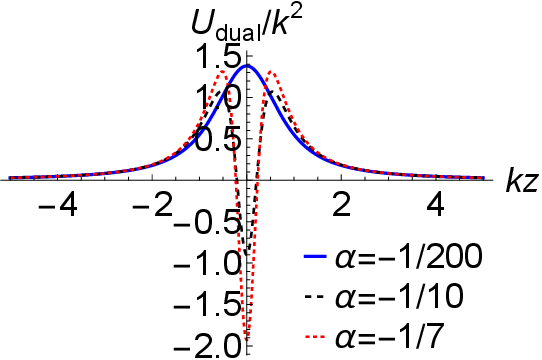}}
	\caption{Plots of the dual potential for different parameters.}
	\label{FigUdual}
\end{figure}

The plots of the effective potential $U_{\text{eff}}(z)$ and the dual potential $U_{\text{dual}}(z)$ for different values of $s$ and $\alpha$ are shown in Figs. \ref{FigUeff} and \ref{FigUdual}, respectively. From Fig. \ref{FigUdual}, we observe that the dual potential develops a double-well structure when $s>1$ or when $\alpha$ becomes sufficiently small. This behavior reflects the splitting of the brane, which was already observed in the energy density plots shown in Figs. \ref{FigRhoS} and \ref{FigRhoAlpha}. Furthermore, when the splitting is induced by $s$, a platform structure appears around near $z=0$, whereas no such platform is present when the splitting is caused by $\alpha$. This feature is consistent with the energy density plots discussed previously in Fig. \ref{FigRho}.    

\par
Since the computation of  QNFs is equivalent to solving the eigenvalue equation \eqref{eq_zdual}, we can apply aforementioned methods to determine the spectrum of QNMs.

\subsection{The asymptotic iteration method and the Bernstein spectral method}
Here we use employ the AIM and the BSM to compute the QNFs of the model. We first briefly review the AIM. In this method, solving an eigenvalue problem is transformed into solving a second-order homogeneous differential equation. Considering the equation

\begin{equation}\label{AIM0}
	y''(x) = \lambda_0(x)y'(x) + s_0(x)y(x),
\end{equation}
where $\lambda_0(x) \neq 0$ and $s_0(x)$ are functions in $C_{\infty}(a,b)$. By differentiating with respect to $x$, one obtains the recurrence relations

\begin{equation}\label{AIM0.5}
	y^{(n+1)}(x) = \lambda_{n-1}(x)y' + s_{n-1}(x)y,
	\quad
	y^{(n+2)}(x) = \lambda_n(x)y' + s_n(x)y,
\end{equation}
where 

\begin{equation}
	\lambda_n = \lambda'_{n-1} + s_{n-1} + \lambda_0\lambda_{n-1},
	\quad
	s_n = s'_{n-1} + s_0\lambda_{n-1}.
\end{equation}
Then we can get

\begin{equation}\label{AIM1}
	\frac{d}{dx} \ln\left(y^{(n+1)}\right) = \frac{y^{(n+2)}}{y^{(n+1)}} = \frac{\lambda_n \left( y'+\frac{s_n}{\lambda_n} \right)}{\lambda_n \left( y'+\frac{s_{n-1}}{\lambda_{n-1}} \right)}.
\end{equation}
If the following condition is satisfied

\begin{equation}\label{AIM2}
	\frac{s_n}{\lambda_n} = \frac{s_{n-1}}{\lambda_{n-1}} \equiv \beta,
\end{equation}
then Eq. \eqref{AIM1} becomes

\begin{equation}\label{AIM2.5}
	\frac{d}{dx} \ln(y^{(n+1)}) = \frac{\lambda_n}{\lambda_{n-1}}.
\end{equation}
The solution of Eq. \eqref{AIM2.5} can be written as

\begin{equation}\label{AIM3}
	y^{(n+1)}(x) = C_1\lambda_{n-1} \exp\left( \int^{x}(\beta + \lambda_0)dt \right).
\end{equation}
Combining Eq. \eqref{AIM3} with Eqs. \eqref{AIM0} and \eqref{AIM2}, we obtain

\begin{equation}
	y' + \beta y = C_1 \exp \left( \int^{x}(\beta + \lambda_0)dt \right),
\end{equation}
whose solution is 

\begin{equation}
	y(x) = \exp\left( -\int^{x}\beta dt \right) \left(C_2 + C_1 \int^{x} \exp \left\{\int^{t}\left[ \lambda_0(\tau)+2\beta(\tau) \right] d\tau \right\} dt \right).
\end{equation} 
The above process shows that a nontrivial solution of Eq. \eqref{AIM0} can be obtained if 

\begin{equation}
	\exists n\in \mathbb{N}, 
	\text{ s.t. }
	\frac{s_n}{\lambda_n} = \frac{s_{n-1}}{\lambda_{n-1}}.
\end{equation}
Expanding $\lambda_n$ and $s_n$ in a Taylor series around the point $\xi$, where the AIM is implemented \cite{AIM_2011}, we obtain

\begin{equation}\label{AIM4}
	\lambda_n(\xi) = \sum^{\infty}_{i=0} c_n^i(x-\xi)^i,
	\qquad
	s_n(\xi) = \sum^{\infty}_{i=0} d_n^i(x-\xi)^i.
\end{equation}
Substituting Eq. \eqref{AIM4} into Eq. \eqref{AIM0.5} yields

\begin{equation}
	c^i_n = (i+1)c^{i+1}_{n-1} + d^i_{n-1} + \Sigma^{i}_{k=0} c^k_0 c^{i-k}_{n-1},
	\qquad
	d^i_n = (i+1)d^{i+1}_{n-1} + \Sigma^i_{k=0} d^k_0 c^{i-k}_{n-1}.
\end{equation}
Therefore, Eq. \eqref{AIM2} can be simplified as

\begin{equation}\label{AIM5}
	d^0_n c^0_{n-1} - d^0_{n-1} c^0_n = 0.
\end{equation}

In this way, derivative operators are no longer required, which significantly reduces the computational complexity. Comparing the eigenvalue equation \eqref{eq_zdual} with the differential equation considered in the AIM, we find that Eq. \eqref{eq_zdual} lacks the first derivative term. Therefore, we perform the coordinate transformation $u = \frac{\sqrt{4k^2z^2+1} - 1 }{2kz}$ \cite{Tan5}. Then Eq. \eqref{eq_zdual} becomes 

\begin{equation}\label{eq_u}
	w_2(u)\hat{\varphi}''(u) + w_1(u)\hat{\varphi}'(u) + w_0(u)\hat{\varphi}(u) = 0,
\end{equation}
where
	
\begin{equation}
	w_2(u) = \frac{k^2 \left(u^2-1\right)^4}{\left(u^2+1\right)^2},
	\qquad
	w_1(u) = \frac{2uk^2 \left(u^2-1\right)^3 \left(u^2+3\right)}{\left(u^2+1\right)^3}.
\end{equation}
For brevity, we set $\alpha = -1/200$, $s = \delta = 1$, and omit the explicit form of $w_0(u)$ due to its excessive length. 
The domain of $\hat{\varphi}(u)$ is now $u\in(-1,1)$, and the boundary conditions become 

\begin{equation}
	\hat{\varphi}(u)\propto
	\begin{cases}
		e^{-\frac{im/k}{2u-2}}, & u\to1,\\
		e^{\frac{im/k}{2u+2}}, & u\to-1.
	\end{cases}
\end{equation}
Thus we write the solution $\hat{\varphi}(u)$ as

\begin{equation}\label{PhiWithBoundary}
	\hat{\varphi}(u) = \hat{\Phi}(u)e^{-\frac{im/k}{2u-2}}e^{\frac{im/k}{2u+2}}.
\end{equation}
Substituting the above equation \eqref{PhiWithBoundary} into Eq. \eqref{eq_u}, we obtain

\begin{equation}\label{eq_u_WithBoundary}
	\hat{\Phi}''(u) = v_1(u)\hat{\Phi}'(u) + v_0(u)\hat{\Phi}(u),
\end{equation}
where
\begin{equation}
	v_1(u) = -\frac{2u \left( 2im \left(u^2+1\right)+u^4+2u^2-3 \right)}{ \left(u^2+1\right) \left(u^2-1\right)^2}.
\end{equation}
Again, $v_0(u)$ is omitted for brevity. The AIM can now be applied to solve the Eq. \eqref{eq_u_WithBoundary}.

The resulting spectrum is shown in Table \ref{AIM_And_BSM}. Note that the obtained quantities correspond to the masses of the KK modes. However, the QNFs $\omega$ can be determined indirectly through the relation $m^2 = \omega^2 - a^2$. Since $a$ is taken to be real, we get $\text{Im}(m) = \text{Im}(\omega)$. Therefore, as $|m|$ increases, the lifetime of KK particles becomes shorter, as can be inferred from Eq. \eqref{PsiToVarphi}.

\begin{table}[htbp]
	\centering
	\begin{tblr}{
			cells = {c},
			cell{1}{2} = {c=2}{},
			cell{1}{4} = {c=2}{},
			vline{1-2,4,6} = {-}{},
			hline{1-2,11} = {-}{},
		}
		$n$ & Asymptotic iteration method &   & Bernstein spectral method &   \\
		&   Re$(m/k)$      &      Im$(m/k)$      &      Re$(m/k)$      &     Im$(m/k)$ \\
		1 &   0.961018 &   -0.504990 &   0.961018 &  -0.504990 \\
		2 &   0.580656 &   -1.78740  &   0.57916  &  -1.79082 \\
		3 &   0.448830 &   -3.44661	&   0.40071	 &  -3.45419 \\
		4 &   0.341653 &   -5.01998	&   0.39192	 &  -5.09895 \\
		5 &   0.593367 &   -9.97493	&   0.58086	 &  -11.5944 \\
		6 &   0.648458 &   -13.1536	&   0.63280  &	-13.2187 \\
		7 &   0.998788 &   -24.6320	&   0.99410	 &  -27.8837 \\
		8 &   1.12214  &   -29.5052	&   1.00620	 &  -29.5395   	
	\end{tblr}
	\caption{$m$ obtained by AIM and BSM for $\alpha=-1/200, \delta=1, s=1$}
	\label{AIM_And_BSM}
\end{table}

As for the BSM, it assumes that the solutions of the differential equation can be expanded in terms of Bernstein polynomials, which serve as a set of basis functions. In this way, the differential equation can be transformed into a system of algebraic equations involving the expansion coefficients. Specifically, we consider the differential equation

\begin{equation}\label{eqBS}
	\hat{L}(u,\omega)\phi(u) = 0,
\end{equation}
where the $\hat{L}(u,\omega)$ is a linear differential operator that depends only on the variable $u$ and eigenvalue $\omega$, and $\phi(u)$ is a function of $u$. We require that the eigenvalue problem satisfies the following conditions: 
\begin{enumerate}
	\item The domain of $\phi(u)$ is compact and analytic over the entire domain, i.e., $u\in [a,b]$.
	\item The boundary behavior of all eigenfunctions $\psi_i(u)$ satisfies $\text{lim}_{u \to a}\psi_i(u) \to (u-a)^q$ and $\text{lim}_{u \to b}\psi_i(u) \to (b-u)^r$ for some $q,r\ge0$.
	\item The eigenvalues of $\omega$ form a discrete spectrum.
\end{enumerate}

We then expand the solution in terms of Bernstein polynomials as 

\begin{equation}
	\phi(u) = \Sigma^N_{k=0}C_kB^N_k(u),
\end{equation}
where $B^N_k(u) = \tbinom{N}{k}\frac{(u-a)^k(b-u)^{N-k}}{(b-a)^N}$ is the Bernstein polynomial, $N$ denotes the order of  Bernstein basis adopted in the expansion. Substituting this expresstion into Eq. \eqref{eqBS}, we obtain a matrix equation,

\begin{equation}
	\mathbf{M}(\omega)\mathbf{C} = 0.
\end{equation}
The above procedure is quite general, but the resulting equation becomes significantly simpler due to the properties of the Bernstein polynomials and the boundary conditions imposed on the eigenfunctions. Since Eq. \eqref{eq_u_WithBoundary} satisfies the conditions required by the BSM, we can apply this method to solve the problem. 

\begin{figure}[htbp]
	\centering
	\includegraphics[width=0.6\textwidth]{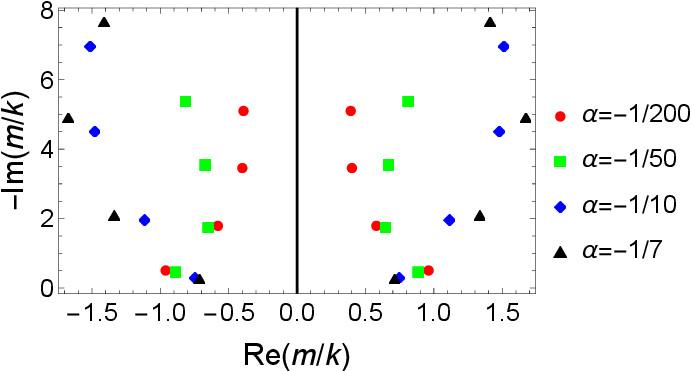}
	\caption{The first four $m$ for different values of $\alpha$ with $s=1$ and $\delta=1$.}
	\label{FigQNMPlotAlpha}
\end{figure}

The obtained results are also presented in Table \ref{AIM_And_BSM}. We find that the results obtained from the AIM and BSM are consistent when the overtone number $n$ is small. However, as $n$ increases, the differences between the results become more noticeable, although the overall trend remains the same. 
To further illustrate the behavior, we plot the spectrum for different values of $\alpha$, as shown in Fig. \ref{FigQNMPlotAlpha}. We find that when $n=1$, the lifetime of the KK particles increases as $|\alpha|$ increases. In contrast, for $n>1$, the overall trend is reversed.

\subsection{Time evolution}

\begin{figure}[htbp]
	\centering
	\subfigure[$\delta=11,s=1,\alpha=-1/200$]{\label{FigEvenA1Delta11Z10}
		\includegraphics[width=0.3\textwidth]{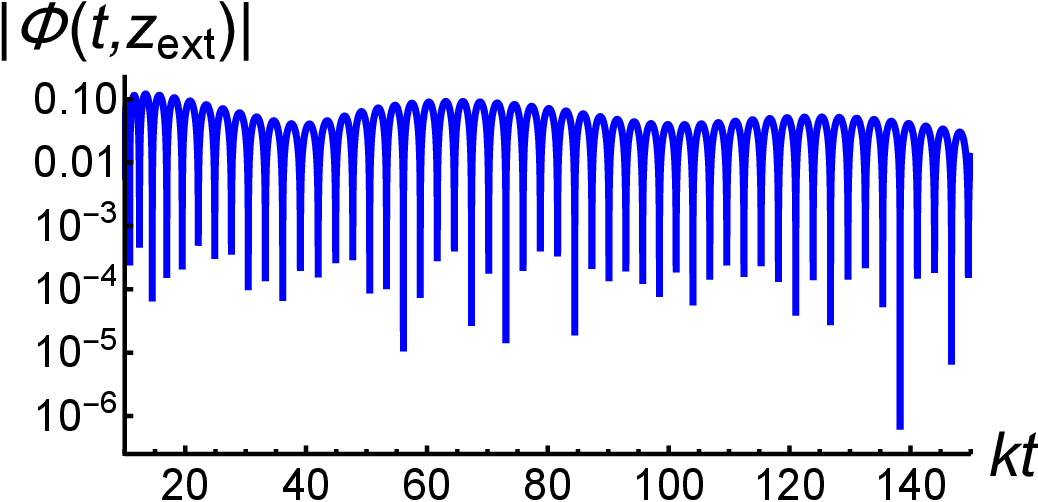}}
	\subfigure[$\delta=1,s=11,\alpha=-1/200$]{\label{FigEvenA1S11Z10}
		\includegraphics[width=0.3\textwidth]{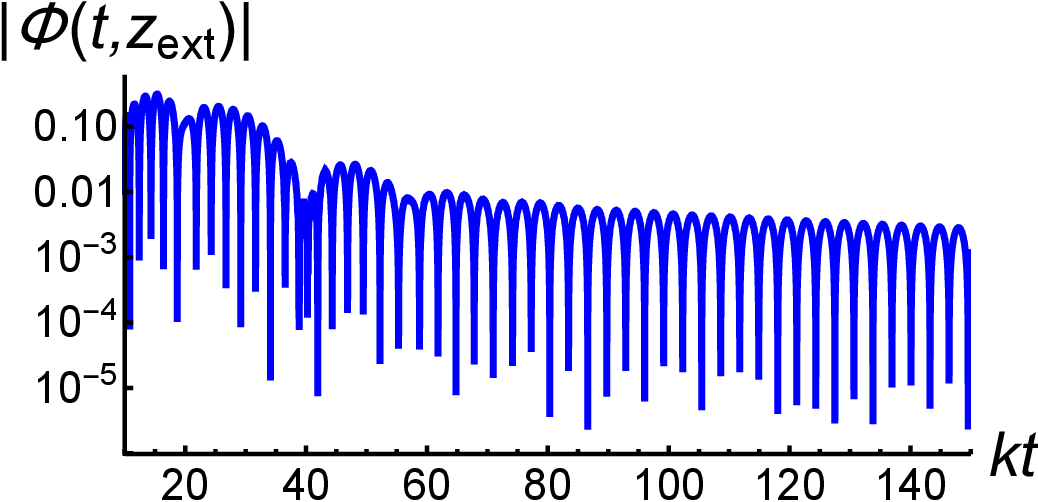}}
	\subfigure[$\delta=1,s=1,\alpha=-1/7$]{\label{FigEvenA1Alpha-17Z2}
		\includegraphics[width=0.3\textwidth]{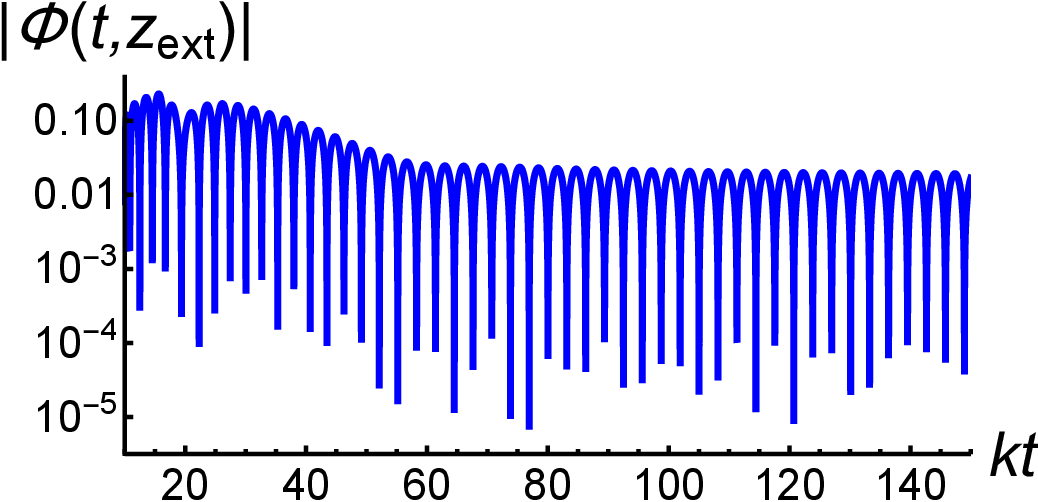}}
	\caption{Evolution waveform of the Gauss wave packet for different parameters at $kz_{\text{ext}}=10$ when $a=1$.}
	\label{FigTimeEvolutionEven}
\end{figure}

\begin{figure}[htbp]
	\centering
	\includegraphics[width=0.4\textwidth]{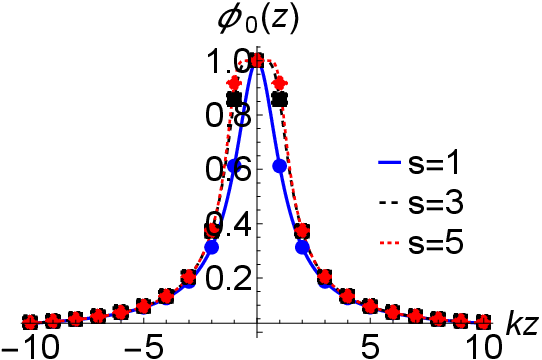}
	\caption{Zero modes (dot) excited by the Gauss pulses and the analytical zero modes (line) for $\alpha=-1/200$ and $\delta =1$.}
	\label{ZeroMode}
\end{figure}

Now we consider the evolution of an initial wave packet incident on the brane. It is convenient to introduce the light-cone coordinates $u = t + z$ and $v = t - z$. Then Eq. \eqref{eq_tz} can be rewritten as 

\begin{equation}
	\left(4 \frac{\partial^2}{\partial u \partial v} + U_{\text{eff}} + a^2 \right)\psi = 0.
\end{equation}
We first consider a Gaussian pulse, 

\begin{equation}
	\psi(0,v) = \exp\left( \frac{-(v-v_c)^2}{2\sigma^2} \right),
	\qquad
	\psi(u,0) = \exp\left( \frac{-v_c^2}{2\sigma^2} \right),
\end{equation}
where $v_c$ determines the center of the wave packet and $\sigma$ determines its width. In the following analysis, we fix $v_c = 5$ and $\sigma = 1$. The numerical evolution for different parameters is shown in Fig. \ref{FigTimeEvolutionEven}, where $z_{\text{ext}}$ denotes the position of the observer in the extra dimension. Although the curves do not differ significantly for different parameter choices, a common feature appears when the Gaussian pulse is incident on the brane: the amplitude of the evolution eventually decays to a constant value. 
When $m = 0$, we can obtain the solution of Eq. \eqref{eq_zeff} which is the zero mode,
\begin{equation}
	\varphi_{0} = N_{0} e^{\frac{3}{2}A-\int K(z)dz},
\end{equation}
where $N_{0}$ is the normalization coefficient. We plot the zero modes for different $s$, as shown in Fig. \ref{ZeroMode}. Correspondingly, the asymptotic constant values of the Gauss wave packets' late-time evolution for different $s$ are extracted, and they are displayed as a scatter plot in Fig. \ref{ZeroMode}, too. From the plot, we can see that the constants are consistent with the zero mode.

\begin{figure}[htbp]
	\centering
	\subfigure[$\delta = 1, s = 1$]{\label{FigTimeEvolutionOddAlpha}
		\includegraphics[width=0.4\textwidth]{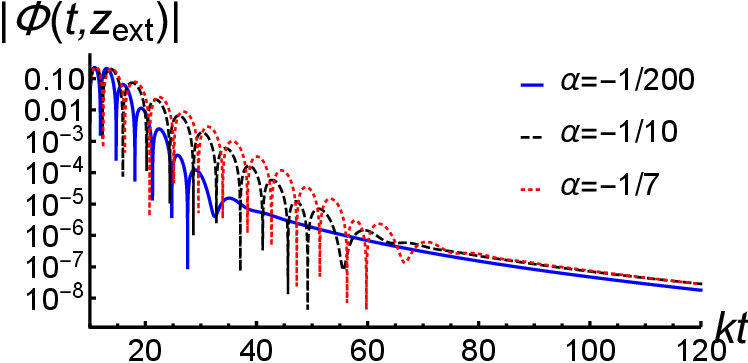}}
	\subfigure[$\delta = 1, \alpha = -1/200$]{\label{FigTimeEvolutionOddS}
		\includegraphics[width=0.4\textwidth]{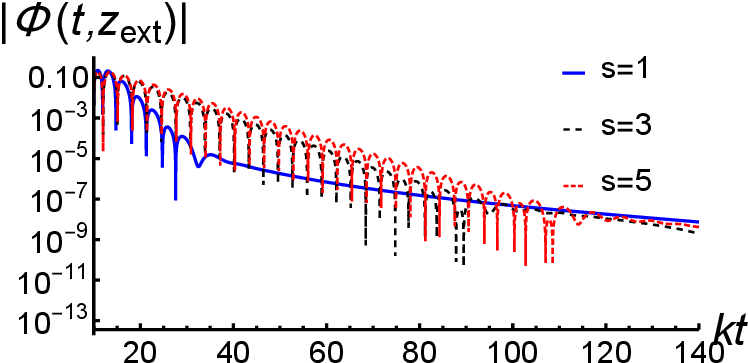}}
	\caption{Evolution waveform of the odd packet with $a=0$ at $kz_{\text{ext}} = 10$ for different parameters in logarithmic scale.}
	\label{FigTimeEvolutionOdd}
\end{figure}

Because the zero mode obscures the quasinormal modes. In order to obtain the information of quasinormal modes, we can consider the case where an odd wave pocket is incident on the brane,

\begin{equation}
	\psi(0,v) = \sin \left(\frac{kv}{2} \right) \exp \left(\frac{-k^2 v^2}{4} \right),
	\qquad
	\psi(u,0) = \sin \left(\frac{ku}{2} \right) \exp \left(\frac{-k^2 u^2}{4} \right).
\end{equation}
The corresponding evolutions for different values of $s$ and $\alpha$ at $kz_{\text{ext}} = 10$ are shown in Fig. \ref{FigTimeEvolutionOdd}. From the figure, we observe that the damping of the wave packet becomes slower as $|\alpha|$ increases, which is consistent with the behavior shown in Fig. \ref{FigQNMPlotAlpha}. A similar tendency is also observed when $s$ increases. Another noteworthy feature appears in Fig. \ref{FigTimeEvolutionOddS}: when $s>1$, the decay rate is significantly reduced. This reflects the fact that the brane undergoes a pronounced splitting when $s>1$, which can also be seen from Fig. \ref{FigRhoS}. Nevertheless, both Figs. \ref{FigTimeEvolutionOddAlpha} and \ref{FigTimeEvolutionOddS} exhibit the characteristic behavior: after an initial exponential decay, the waveform develops a power-law tail.

It is also instructive to compare the cases of odd and even incident wave packets. This difference originates from the wave equation \eqref{eq_tz} and the values of $a$ chosen in our analysis, which satisfy $m^2 = \omega^2 - a^2$. 
The solutions of Eq. \eqref{eq_tz} can be classified into two families according to parity: odd-parity solutions and even-parity solutions. The parity of the incident wave packet therefore determines which class of solutions is excited. Since the zero mode is even parity, it is not excited when an odd wave packet is incident. Furthermore, when $a = 0$, we obtain $m = \omega$. In this case the massless zero mode is effectively removed from the spectrum. Therefore when an odd wave packet is considered together with $a = 0$, the zero mode is completely excluded, and the waveform clearly exhibits the typical behavior mentioned above. Under these conditions, the first QNF can be extracted by fitting the numerical waveform, which can be verified by the DIM. An example is presented in Table \ref{TE_And_DIM} where shows $m$ for different $s$ obtained by the numerical evolution and the DIM. From the results we find that the lifetime of the KK particles increases as $s$ increases, which is consistent with the behavior shown in Fig. \ref{FigTimeEvolutionOddS}. 
In contrast, when a Gaussian wave packet is incident, the zero mode is excited. After some time, the zero mode dominates the signal and masks the other QNMs and the tail. Because the zero mode is localized on the brane and does not decay, the amplitude of the wave packet eventually approaches a constant value. 

\begin{figure}[htbp]
	\centering
	\subfigure[$\delta = 3$]{\label{FigBeatDelta3}
		\includegraphics[width=0.4\textwidth]{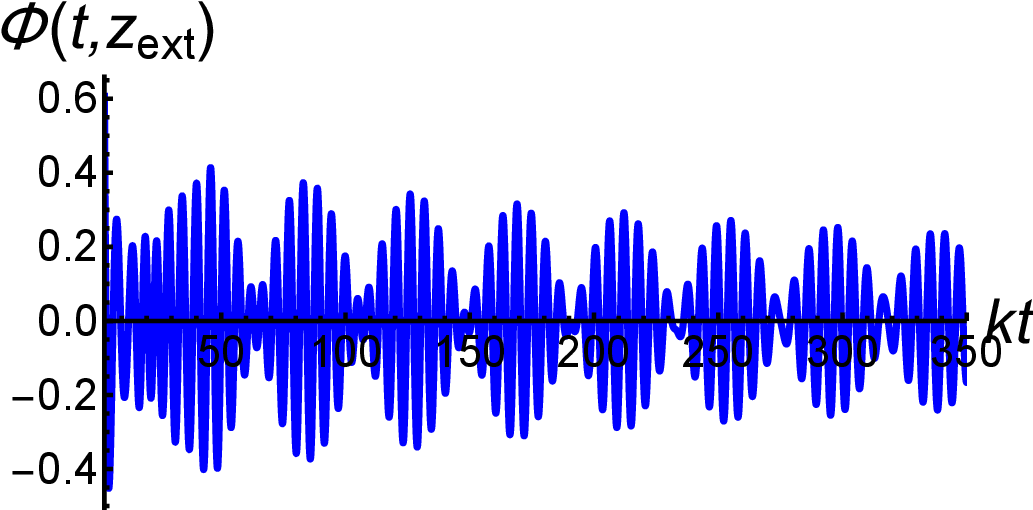}}
	\subfigure[$\delta = 5$]{\label{FigBeatDelta5}
		\includegraphics[width=0.4\textwidth]{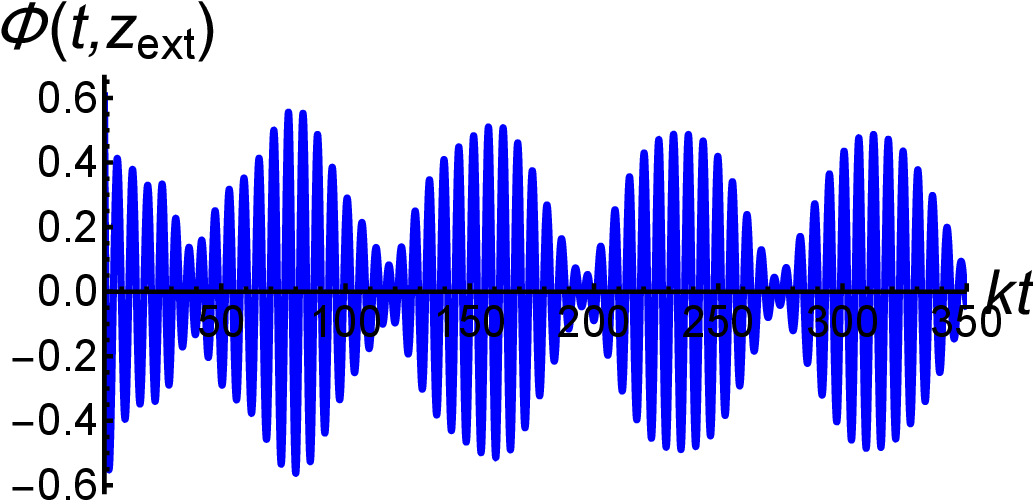}}
	\caption{Beat effect when a Gauss wave packet is incident with $a=1$, $s=3$ and $\alpha=-1/200$ at $kz_{\text{ext}} = 3$.}
	\label{FigBeat}
\end{figure}

Finally, we also observe a beat phenomenon when $\delta$ and $s$ are sufficiently large, as shown in Fig. \ref{FigBeat}. This beat phenomenon is frequently encountered in practical situations, such as the vibrating tuning fork, and beats arise from the superposition of waves with nearly equal frequencies, the same vibration direction, and low damping. This indicates that, in such cases, there exist even-parity modes with very close frequencies and a sufficiently long lifetime leading to the interference pattern observed in the waveform \cite{Tan1}. 

\begin{table}[htbp]
	\centering
	\begin{tblr}{
			cells = {c},
			cell{1}{2} = {c=2}{},
			cell{1}{4} = {c=2}{},
			vline{1-2,4,6} = {-}{},
			hline{1-2,6} = {-}{},
		}
		$s$ & Time evolution &   & Direct Integral method &   \\
		&   Re$(m/k)$      &      Im$(m/k)$      &      Re$(m/k)$      &     Im$(m/k)$ \\
		$1$ &   0.968788  &   -0.528021  &   0.961018 &  -0.504990 \\
		$3$ &   1.00285 &   -0.194022  &   1.00228   &  -0.193428 \\
		$5$ &   0.999519 &   -0.165954  &   0.999300 &  -0.165679 \\   	
	\end{tblr}
	\caption{$m$ obtained by the time evolution and the DIM for $\alpha=-1/200, \delta=1$ and different $s$.}
	\label{TE_And_DIM}
\end{table}

\section{Conclusion}
\label{SecCon}

In this paper, we studied the QNMs of the thick brane model in $f(T) = T + \alpha T^2$ gravity. We found that, in order to ensure the energy density is positive and the scalar field is a real field, the value range of $\alpha$ should be [$-\frac{7}{48}$,$\frac{1}{48}$] in this model. With asymptotic iteration method and Bernstein spectral method, we found in $f(T)$ gravity, there still is a zero mode and a series of discrete QNMs which is similar to the result of the model in general gravity which was studied in \cite{Tan1}. This is predictable because the range of values for  $|\alpha|$ is so small. When $s$ is lager than 1, the brane will split and form a platform near $z=0$. This will extend the lifespan of KK particles significantly. $\alpha$ will also lead to brane splitting while it will not form a platform. When $\alpha<0$, the decay rate of the first QNM decreases but the decay rate of other QNMs increases as $\alpha$ decreases. We extract the waveform of the zero mode from the evolution caused by Gauss wave packet and it is consistent with the analytical solution. This indicates that when an even parity wave packet is incident, the even parity mode will be excited which includes the zero mode and then in the late evolutionary period, the zero mode becomes dominant.

There are many ways to strengthen our work. First, when $s>1$, the QNFs is hard to get through the above methods. We need to find other new ways to calculate. Then, the different models of $f(T)$ gravity and different warp factors could be studied in the future. 

\section{Acknowledgments}
This work was supported by 
the National Natural Science Foundation of China (Grants No. 12205129 and No. 12247101),
the Fundamental Research Funds for the Central Universities (Grants No. lzujbky-2025-it05 and lzujbky-2025-jdzx07), 
the Natural Science Foundation of Gansu Province (No. 22JR5RA389, No.25JRRA799), 
the ‘111 Center’ under Grant No. B20063,
and the Key Project of the Department of Education of Hunan Province (No. 25A0084).
Wen-Di Guo was supported by “Talent Scientific Fund of Lanzhou University”.

\nocite{*}

\FloatBarrier

\bibliography{f_T_QNM}

\providecommand{\noopsort}[1]{}\providecommand{\singleletter}[1]{#1}%
\begin{thebibliography}{82}%
\makeatletter
\providecommand \@ifxundefined [1]{%
 \@ifx{#1\undefined}
}%
\providecommand \@ifnum [1]{%
 \ifnum #1\expandafter \@firstoftwo
 \else \expandafter \@secondoftwo
 \fi
}%
\providecommand \@ifx [1]{%
 \ifx #1\expandafter \@firstoftwo
 \else \expandafter \@secondoftwo
 \fi
}%
\providecommand \natexlab [1]{#1}%
\providecommand \enquote  [1]{``#1''}%
\providecommand \bibnamefont  [1]{#1}%
\providecommand \bibfnamefont [1]{#1}%
\providecommand \citenamefont [1]{#1}%
\providecommand \href@noop [0]{\@secondoftwo}%
\providecommand \href [0]{\begingroup \@sanitize@url \@href}%
\providecommand \@href[1]{\@@startlink{#1}\@@href}%
\providecommand \@@href[1]{\endgroup#1\@@endlink}%
\providecommand \@sanitize@url [0]{\catcode `\\12\catcode `\$12\catcode
  `\&12\catcode `\#12\catcode `\^12\catcode `\_12\catcode `\%12\relax}%
\providecommand \@@startlink[1]{}%
\providecommand \@@endlink[0]{}%
\providecommand \url  [0]{\begingroup\@sanitize@url \@url }%
\providecommand \@url [1]{\endgroup\@href {#1}{\urlprefix }}%
\providecommand \urlprefix  [0]{URL }%
\providecommand \Eprint [0]{\href }%
\providecommand \doibase [0]{https://doi.org/}%
\providecommand \selectlanguage [0]{\@gobble}%
\providecommand \bibinfo  [0]{\@secondoftwo}%
\providecommand \bibfield  [0]{\@secondoftwo}%
\providecommand \translation [1]{[#1]}%
\providecommand \BibitemOpen [0]{}%
\providecommand \bibitemStop [0]{}%
\providecommand \bibitemNoStop [0]{.\EOS\space}%
\providecommand \EOS [0]{\spacefactor3000\relax}%
\providecommand \BibitemShut  [1]{\csname bibitem#1\endcsname}%
\let\auto@bib@innerbib\@empty
\bibitem [{\citenamefont {Nordstrom}(1914)}]{Nordstrom}%
  \BibitemOpen
  \bibfield  {author} {\bibinfo {author} {\bibfnamefont {G.}~\bibnamefont
  {Nordstrom}},\ }\bibfield  {title} {\bibinfo {title} {{On the possibility of
  unifying the electromagnetic and the gravitational fields}},\ }\href@noop {}
  {\bibfield  {journal} {\bibinfo  {journal} {Phys. Z.}\ }\textbf {\bibinfo
  {volume} {15}},\ \bibinfo {pages} {504} (\bibinfo {year} {1914})},\ \Eprint
  {https://arxiv.org/abs/physics/0702221} {arXiv:physics/0702221} \BibitemShut
  {NoStop}%
\bibitem [{\citenamefont {Kaluza}(1921)}]{Kaluza}%
  \BibitemOpen
  \bibfield  {author} {\bibinfo {author} {\bibfnamefont {T.}~\bibnamefont
  {Kaluza}},\ }\bibfield  {title} {\bibinfo {title} {Zum unit{\"a}tsproblem der
  physik},\ }\href@noop {} {\bibfield  {journal} {\bibinfo  {journal}
  {Sitzungsber.\ Preuss.\ Akad.\ Wiss.\ Berlin\ (Math.\ Phys. )}\ }\textbf
  {\bibinfo {volume} {1921}},\ \bibinfo {pages} {966} (\bibinfo {year}
  {1921})},\ \Eprint {https://arxiv.org/abs/1803.08616} {arXiv:1803.08616
  [physics.hist-ph]} \BibitemShut {NoStop}%
\bibitem [{\citenamefont {Klein}(1926)}]{Klein}%
  \BibitemOpen
  \bibfield  {author} {\bibinfo {author} {\bibfnamefont {O.}~\bibnamefont
  {Klein}},\ }\bibfield  {title} {\bibinfo {title} {{Quantum Theory and
  Five-Dimensional Theory of Relativity. (In German and English)}},\ }\href
  {https://doi.org/10.1007/BF01397481} {\bibfield  {journal} {\bibinfo
  {journal} {Z. Phys.}\ }\textbf {\bibinfo {volume} {37}},\ \bibinfo {pages}
  {895} (\bibinfo {year} {1926})}\BibitemShut {NoStop}%
\bibitem [{\citenamefont {Akama}(1982)}]{Akama}%
  \BibitemOpen
  \bibfield  {author} {\bibinfo {author} {\bibfnamefont {K.}~\bibnamefont
  {Akama}},\ }\bibfield  {title} {\bibinfo {title} {{An Early Proposal of
  ``Brane World''}},\ }\href@noop {} {\bibfield  {journal} {\bibinfo  {journal}
  {Lect. Notes Phys.}\ }\textbf {\bibinfo {volume} {176}},\ \bibinfo {pages}
  {267} (\bibinfo {year} {1982})},\ \Eprint
  {https://arxiv.org/abs/hep-th/0001113} {arXiv:hep-th/0001113} \BibitemShut
  {NoStop}%
\bibitem [{\citenamefont {Arkani-Hamed}\ \emph {et~al.}(1998)\citenamefont
  {Arkani-Hamed}, \citenamefont {Dimopoulos},\ and\ \citenamefont
  {Dvali}}]{ADD}%
  \BibitemOpen
  \bibfield  {author} {\bibinfo {author} {\bibfnamefont {N.}~\bibnamefont
  {Arkani-Hamed}}, \bibinfo {author} {\bibfnamefont {S.}~\bibnamefont
  {Dimopoulos}},\ and\ \bibinfo {author} {\bibfnamefont {G.~R.}\ \bibnamefont
  {Dvali}},\ }\bibfield  {title} {\bibinfo {title} {{The Hierarchy problem and
  new dimensions at a millimeter}},\ }\href
  {https://doi.org/10.1016/S0370-2693(98)00466-3} {\bibfield  {journal}
  {\bibinfo  {journal} {Phys. Lett. B}\ }\textbf {\bibinfo {volume} {429}},\
  \bibinfo {pages} {263} (\bibinfo {year} {1998})},\ \Eprint
  {https://arxiv.org/abs/hep-ph/9803315} {arXiv:hep-ph/9803315} \BibitemShut
  {NoStop}%
\bibitem [{\citenamefont {Randall}\ and\ \citenamefont
  {Sundrum}(1999{\natexlab{a}})}]{Randall1}%
  \BibitemOpen
  \bibfield  {author} {\bibinfo {author} {\bibfnamefont {L.}~\bibnamefont
  {Randall}}\ and\ \bibinfo {author} {\bibfnamefont {R.}~\bibnamefont
  {Sundrum}},\ }\bibfield  {title} {\bibinfo {title} {{A Large mass hierarchy
  from a small extra dimension}},\ }\href
  {https://doi.org/10.1103/PhysRevLett.83.3370} {\bibfield  {journal} {\bibinfo
   {journal} {Phys. Rev. Lett.}\ }\textbf {\bibinfo {volume} {83}},\ \bibinfo
  {pages} {3370} (\bibinfo {year} {1999}{\natexlab{a}})},\ \Eprint
  {https://arxiv.org/abs/hep-ph/9905221} {arXiv:hep-ph/9905221} \BibitemShut
  {NoStop}%
\bibitem [{\citenamefont {Randall}\ and\ \citenamefont
  {Sundrum}(1999{\natexlab{b}})}]{Randall2}%
  \BibitemOpen
  \bibfield  {author} {\bibinfo {author} {\bibfnamefont {L.}~\bibnamefont
  {Randall}}\ and\ \bibinfo {author} {\bibfnamefont {R.}~\bibnamefont
  {Sundrum}},\ }\bibfield  {title} {\bibinfo {title} {{An Alternative to
  compactification}},\ }\href {https://doi.org/10.1103/PhysRevLett.83.4690}
  {\bibfield  {journal} {\bibinfo  {journal} {Phys. Rev. Lett.}\ }\textbf
  {\bibinfo {volume} {83}},\ \bibinfo {pages} {4690} (\bibinfo {year}
  {1999}{\natexlab{b}})},\ \Eprint {https://arxiv.org/abs/hep-th/9906064}
  {arXiv:hep-th/9906064} \BibitemShut {NoStop}%
\bibitem [{\citenamefont {Shiromizu}\ \emph {et~al.}(2000)\citenamefont
  {Shiromizu}, \citenamefont {Maeda},\ and\ \citenamefont
  {Sasaki}}]{Shiromizu}%
  \BibitemOpen
  \bibfield  {author} {\bibinfo {author} {\bibfnamefont {T.}~\bibnamefont
  {Shiromizu}}, \bibinfo {author} {\bibfnamefont {K.-i.}\ \bibnamefont
  {Maeda}},\ and\ \bibinfo {author} {\bibfnamefont {M.}~\bibnamefont
  {Sasaki}},\ }\bibfield  {title} {\bibinfo {title} {{The Einstein equation on
  the 3-brane world}},\ }\href {https://doi.org/10.1103/PhysRevD.62.024012}
  {\bibfield  {journal} {\bibinfo  {journal} {Phys. Rev. D}\ }\textbf {\bibinfo
  {volume} {62}},\ \bibinfo {pages} {024012} (\bibinfo {year} {2000})},\
  \Eprint {https://arxiv.org/abs/gr-qc/9910076} {arXiv:gr-qc/9910076}
  \BibitemShut {NoStop}%
\bibitem [{\citenamefont {Tanaka}(2003)}]{Tanaka}%
  \BibitemOpen
  \bibfield  {author} {\bibinfo {author} {\bibfnamefont {T.}~\bibnamefont
  {Tanaka}},\ }\bibfield  {title} {\bibinfo {title} {{Classical black hole
  evaporation in Randall-Sundrum infinite brane world}},\ }\href
  {https://doi.org/10.1143/PTPS.148.307} {\bibfield  {journal} {\bibinfo
  {journal} {Prog. Theor. Phys. Suppl.}\ }\textbf {\bibinfo {volume} {148}},\
  \bibinfo {pages} {307} (\bibinfo {year} {2003})},\ \Eprint
  {https://arxiv.org/abs/gr-qc/0203082} {arXiv:gr-qc/0203082} \BibitemShut
  {NoStop}%
\bibitem [{\citenamefont {Gregory}(2009)}]{Gregory}%
  \BibitemOpen
  \bibfield  {author} {\bibinfo {author} {\bibfnamefont {R.}~\bibnamefont
  {Gregory}},\ }\bibfield  {title} {\bibinfo {title} {{Braneworld black
  holes}},\ }\href {https://doi.org/10.1007/978-3-540-88460-6_7} {\bibfield
  {journal} {\bibinfo  {journal} {Lect. Notes Phys.}\ }\textbf {\bibinfo
  {volume} {769}},\ \bibinfo {pages} {259} (\bibinfo {year} {2009})},\ \Eprint
  {https://arxiv.org/abs/0804.2595} {arXiv:0804.2595 [hep-th]} \BibitemShut
  {NoStop}%
\bibitem [{\citenamefont {Jaman}\ and\ \citenamefont
  {Myrzakulov}(2019)}]{Jaman}%
  \BibitemOpen
  \bibfield  {author} {\bibinfo {author} {\bibfnamefont {N.}~\bibnamefont
  {Jaman}}\ and\ \bibinfo {author} {\bibfnamefont {K.}~\bibnamefont
  {Myrzakulov}},\ }\bibfield  {title} {\bibinfo {title} {{Braneworld inflation
  with an effective $\alpha$-attractor potential}},\ }\href
  {https://doi.org/10.1103/PhysRevD.99.103523} {\bibfield  {journal} {\bibinfo
  {journal} {Phys. Rev. D}\ }\textbf {\bibinfo {volume} {99}},\ \bibinfo
  {pages} {103523} (\bibinfo {year} {2019})},\ \Eprint
  {https://arxiv.org/abs/1807.07443} {arXiv:1807.07443 [gr-qc]} \BibitemShut
  {NoStop}%
\bibitem [{\citenamefont {Adhikari}\ \emph {et~al.}(2020)\citenamefont
  {Adhikari}, \citenamefont {Gangopadhyay},\ and\ \citenamefont
  {Yogesh}}]{Adhikari}%
  \BibitemOpen
  \bibfield  {author} {\bibinfo {author} {\bibfnamefont {R.}~\bibnamefont
  {Adhikari}}, \bibinfo {author} {\bibfnamefont {M.~R.}\ \bibnamefont
  {Gangopadhyay}},\ and\ \bibinfo {author} {\bibnamefont {Yogesh}},\ }\bibfield
   {title} {\bibinfo {title} {{Power Law Plateau Inflation Potential In The
  RS-\uppercase\expandafter{\romannumeral2} Braneworld Evading Swampland
  Conjecture}},\ }\href {https://doi.org/10.1140/epjc/s10052-020-08460-3}
  {\bibfield  {journal} {\bibinfo  {journal} {Eur. Phys. J. C}\ }\textbf
  {\bibinfo {volume} {80}},\ \bibinfo {pages} {899} (\bibinfo {year} {2020})},\
  \Eprint {https://arxiv.org/abs/2002.07061} {arXiv:2002.07061 [astro-ph.CO]}
  \BibitemShut {NoStop}%
\bibitem [{\citenamefont {Geng}\ \emph
  {et~al.}(2021{\natexlab{a}})\citenamefont {Geng}, \citenamefont {Karch},
  \citenamefont {Perez-Pardavila}, \citenamefont {Raju}, \citenamefont
  {Randall}, \citenamefont {Riojas},\ and\ \citenamefont
  {Shashi}}]{Geng:2020fxl}%
  \BibitemOpen
  \bibfield  {author} {\bibinfo {author} {\bibfnamefont {H.}~\bibnamefont
  {Geng}}, \bibinfo {author} {\bibfnamefont {A.}~\bibnamefont {Karch}},
  \bibinfo {author} {\bibfnamefont {C.}~\bibnamefont {Perez-Pardavila}},
  \bibinfo {author} {\bibfnamefont {S.}~\bibnamefont {Raju}}, \bibinfo {author}
  {\bibfnamefont {L.}~\bibnamefont {Randall}}, \bibinfo {author} {\bibfnamefont
  {M.}~\bibnamefont {Riojas}},\ and\ \bibinfo {author} {\bibfnamefont
  {S.}~\bibnamefont {Shashi}},\ }\bibfield  {title} {\bibinfo {title}
  {{Information Transfer with a Gravitating Bath}},\ }\href
  {https://doi.org/10.21468/SciPostPhys.10.5.103} {\bibfield  {journal}
  {\bibinfo  {journal} {SciPost Phys.}\ }\textbf {\bibinfo {volume} {10}},\
  \bibinfo {pages} {103} (\bibinfo {year} {2021}{\natexlab{a}})},\ \Eprint
  {https://arxiv.org/abs/2012.04671} {arXiv:2012.04671 [hep-th]} \BibitemShut
  {NoStop}%
\bibitem [{\citenamefont {Geng}\ \emph
  {et~al.}(2021{\natexlab{b}})\citenamefont {Geng}, \citenamefont {L{\"u}st},
  \citenamefont {Mishra},\ and\ \citenamefont {Wakeham}}]{Geng:2021iyq}%
  \BibitemOpen
  \bibfield  {author} {\bibinfo {author} {\bibfnamefont {H.}~\bibnamefont
  {Geng}}, \bibinfo {author} {\bibfnamefont {S.}~\bibnamefont {L{\"u}st}},
  \bibinfo {author} {\bibfnamefont {R.~K.}\ \bibnamefont {Mishra}},\ and\
  \bibinfo {author} {\bibfnamefont {D.}~\bibnamefont {Wakeham}},\ }\bibfield
  {title} {\bibinfo {title} {{Holographic BCFTs and Communicating Black
  Holes}},\ }\href {https://doi.org/10.1007/JHEP08(2021)003} {\bibfield
  {journal} {\bibinfo  {journal} {JHEP}\ }\textbf {\bibinfo {volume} {08}},\
  \bibinfo {pages} {003}},\ \Eprint {https://arxiv.org/abs/2104.07039}
  {arXiv:2104.07039 [hep-th]} \BibitemShut {NoStop}%
\bibitem [{\citenamefont {Geng}\ \emph {et~al.}(2022)\citenamefont {Geng},
  \citenamefont {Randall},\ and\ \citenamefont {Swanson}}]{Geng:2022dua}%
  \BibitemOpen
  \bibfield  {author} {\bibinfo {author} {\bibfnamefont {H.}~\bibnamefont
  {Geng}}, \bibinfo {author} {\bibfnamefont {L.}~\bibnamefont {Randall}},\ and\
  \bibinfo {author} {\bibfnamefont {E.}~\bibnamefont {Swanson}},\ }\bibfield
  {title} {\bibinfo {title} {{BCFT in a black hole background: an analytical
  holographic model}},\ }\href {https://doi.org/10.1007/JHEP12(2022)056}
  {\bibfield  {journal} {\bibinfo  {journal} {JHEP}\ }\textbf {\bibinfo
  {volume} {12}},\ \bibinfo {pages} {056}},\ \Eprint
  {https://arxiv.org/abs/2209.02074} {arXiv:2209.02074 [hep-th]} \BibitemShut
  {NoStop}%
\bibitem [{\citenamefont {Bhattacharya}\ \emph {et~al.}(2021)\citenamefont
  {Bhattacharya}, \citenamefont {Bhattacharyya}, \citenamefont {Nandy},\ and\
  \citenamefont {Patra}}]{Bhattacharya:2021jrn}%
  \BibitemOpen
  \bibfield  {author} {\bibinfo {author} {\bibfnamefont {A.}~\bibnamefont
  {Bhattacharya}}, \bibinfo {author} {\bibfnamefont {A.}~\bibnamefont
  {Bhattacharyya}}, \bibinfo {author} {\bibfnamefont {P.}~\bibnamefont
  {Nandy}},\ and\ \bibinfo {author} {\bibfnamefont {A.~K.}\ \bibnamefont
  {Patra}},\ }\bibfield  {title} {\bibinfo {title} {{Islands and complexity of
  eternal black hole and radiation subsystems for a doubly holographic
  model}},\ }\href {https://doi.org/10.1007/JHEP05(2021)135} {\bibfield
  {journal} {\bibinfo  {journal} {JHEP}\ }\textbf {\bibinfo {volume} {05}},\
  \bibinfo {pages} {135}},\ \Eprint {https://arxiv.org/abs/2103.15852}
  {arXiv:2103.15852 [hep-th]} \BibitemShut {NoStop}%
\bibitem [{\citenamefont {Rubakov}\ and\ \citenamefont
  {Shaposhnikov}(1983)}]{Rubakov:1983kt}%
  \BibitemOpen
  \bibfield  {author} {\bibinfo {author} {\bibfnamefont {V.~A.}\ \bibnamefont
  {Rubakov}}\ and\ \bibinfo {author} {\bibfnamefont {M.~E.}\ \bibnamefont
  {Shaposhnikov}},\ }\bibfield  {title} {\bibinfo {title} {{Do we live inside a
  domain wall?}},\ }\href {https://doi.org/10.1016/0370-2693(83)91253-4}
  {\bibfield  {journal} {\bibinfo  {journal} {Phys. Lett. B}\ }\textbf
  {\bibinfo {volume} {125}},\ \bibinfo {pages} {136} (\bibinfo {year}
  {1983})}\BibitemShut {NoStop}%
\bibitem [{\citenamefont {DeWolfe}\ \emph {et~al.}(2000)\citenamefont
  {DeWolfe}, \citenamefont {Freedman}, \citenamefont {Gubser},\ and\
  \citenamefont {Karch}}]{DeWolfe:1999cp}%
  \BibitemOpen
  \bibfield  {author} {\bibinfo {author} {\bibfnamefont {O.}~\bibnamefont
  {DeWolfe}}, \bibinfo {author} {\bibfnamefont {D.~Z.}\ \bibnamefont
  {Freedman}}, \bibinfo {author} {\bibfnamefont {S.~S.}\ \bibnamefont
  {Gubser}},\ and\ \bibinfo {author} {\bibfnamefont {A.}~\bibnamefont
  {Karch}},\ }\bibfield  {title} {\bibinfo {title} {{Modeling the
  fifth-dimension with scalars and gravity}},\ }\href
  {https://doi.org/10.1103/PhysRevD.62.046008} {\bibfield  {journal} {\bibinfo
  {journal} {Phys. Rev. D}\ }\textbf {\bibinfo {volume} {62}},\ \bibinfo
  {pages} {046008} (\bibinfo {year} {2000})},\ \Eprint
  {https://arxiv.org/abs/hep-th/9909134} {arXiv:hep-th/9909134} \BibitemShut
  {NoStop}%
\bibitem [{\citenamefont {Gremm}(2000)}]{Gremm:1999pj}%
  \BibitemOpen
  \bibfield  {author} {\bibinfo {author} {\bibfnamefont {M.}~\bibnamefont
  {Gremm}},\ }\bibfield  {title} {\bibinfo {title} {{Four-dimensional gravity
  on a thick domain wall}},\ }\href
  {https://doi.org/10.1016/S0370-2693(00)00303-8} {\bibfield  {journal}
  {\bibinfo  {journal} {Phys. Lett. B}\ }\textbf {\bibinfo {volume} {478}},\
  \bibinfo {pages} {434} (\bibinfo {year} {2000})},\ \Eprint
  {https://arxiv.org/abs/hep-th/9912060} {arXiv:hep-th/9912060} \BibitemShut
  {NoStop}%
\bibitem [{\citenamefont {Csaki}\ \emph {et~al.}(2000)\citenamefont {Csaki},
  \citenamefont {Erlich}, \citenamefont {Hollowood},\ and\ \citenamefont
  {Shirman}}]{Csaki:2000fc}%
  \BibitemOpen
  \bibfield  {author} {\bibinfo {author} {\bibfnamefont {C.}~\bibnamefont
  {Csaki}}, \bibinfo {author} {\bibfnamefont {J.}~\bibnamefont {Erlich}},
  \bibinfo {author} {\bibfnamefont {T.~J.}\ \bibnamefont {Hollowood}},\ and\
  \bibinfo {author} {\bibfnamefont {Y.}~\bibnamefont {Shirman}},\ }\bibfield
  {title} {\bibinfo {title} {{Universal aspects of gravity localized on thick
  branes}},\ }\href {https://doi.org/10.1016/S0550-3213(00)00271-6} {\bibfield
  {journal} {\bibinfo  {journal} {Nucl. Phys. B}\ }\textbf {\bibinfo {volume}
  {581}},\ \bibinfo {pages} {309} (\bibinfo {year} {2000})},\ \Eprint
  {https://arxiv.org/abs/hep-th/0001033} {arXiv:hep-th/0001033} \BibitemShut
  {NoStop}%
\bibitem [{\citenamefont {Dzhunushaliev}\ and\ \citenamefont
  {Folomeev}(2011)}]{Dzhunushaliev}%
  \BibitemOpen
  \bibfield  {author} {\bibinfo {author} {\bibfnamefont {V.}~\bibnamefont
  {Dzhunushaliev}}\ and\ \bibinfo {author} {\bibfnamefont {V.}~\bibnamefont
  {Folomeev}},\ }\bibfield  {title} {\bibinfo {title} {{Spinor brane}},\ }\href
  {https://doi.org/10.1007/s10714-010-1105-2} {\bibfield  {journal} {\bibinfo
  {journal} {Gen. Rel. Grav.}\ }\textbf {\bibinfo {volume} {43}},\ \bibinfo
  {pages} {1253} (\bibinfo {year} {2011})},\ \Eprint
  {https://arxiv.org/abs/0909.2741} {arXiv:0909.2741 [gr-qc]} \BibitemShut
  {NoStop}%
\bibitem [{\citenamefont {Dzhunushaliev}\ and\ \citenamefont
  {Folomeev}(2012)}]{Dzhunushaliev:2011mm}%
  \BibitemOpen
  \bibfield  {author} {\bibinfo {author} {\bibfnamefont {V.}~\bibnamefont
  {Dzhunushaliev}}\ and\ \bibinfo {author} {\bibfnamefont {V.}~\bibnamefont
  {Folomeev}},\ }\bibfield  {title} {\bibinfo {title} {{Thick brane solutions
  supported by two spinor fields}},\ }\href
  {https://doi.org/10.1007/s10714-011-1276-5} {\bibfield  {journal} {\bibinfo
  {journal} {Gen. Rel. Grav.}\ }\textbf {\bibinfo {volume} {44}},\ \bibinfo
  {pages} {253} (\bibinfo {year} {2012})},\ \Eprint
  {https://arxiv.org/abs/1104.2733} {arXiv:1104.2733 [gr-qc]} \BibitemShut
  {NoStop}%
\bibitem [{\citenamefont {Afonso}\ \emph {et~al.}(2007)\citenamefont {Afonso},
  \citenamefont {Bazeia}, \citenamefont {Menezes},\ and\ \citenamefont
  {Petrov}}]{Afonso:2007gc}%
  \BibitemOpen
  \bibfield  {author} {\bibinfo {author} {\bibfnamefont {V.~I.}\ \bibnamefont
  {Afonso}}, \bibinfo {author} {\bibfnamefont {D.}~\bibnamefont {Bazeia}},
  \bibinfo {author} {\bibfnamefont {R.}~\bibnamefont {Menezes}},\ and\ \bibinfo
  {author} {\bibfnamefont {A.~Y.}\ \bibnamefont {Petrov}},\ }\bibfield  {title}
  {\bibinfo {title} {{$f(R)$-Brane}},\ }\href
  {https://doi.org/10.1016/j.physletb.2007.10.038} {\bibfield  {journal}
  {\bibinfo  {journal} {Phys. Lett. B}\ }\textbf {\bibinfo {volume} {658}},\
  \bibinfo {pages} {71} (\bibinfo {year} {2007})},\ \Eprint
  {https://arxiv.org/abs/0710.3790} {arXiv:0710.3790 [hep-th]} \BibitemShut
  {NoStop}%
\bibitem [{\citenamefont {Liu}\ \emph {et~al.}(2011)\citenamefont {Liu},
  \citenamefont {Zhong}, \citenamefont {Zhao},\ and\ \citenamefont
  {Li}}]{Liu:2011wi}%
  \BibitemOpen
  \bibfield  {author} {\bibinfo {author} {\bibfnamefont {Y.-X.}\ \bibnamefont
  {Liu}}, \bibinfo {author} {\bibfnamefont {Y.}~\bibnamefont {Zhong}}, \bibinfo
  {author} {\bibfnamefont {Z.-H.}\ \bibnamefont {Zhao}},\ and\ \bibinfo
  {author} {\bibfnamefont {H.-T.}\ \bibnamefont {Li}},\ }\bibfield  {title}
  {\bibinfo {title} {{Domain wall brane in squared curvature gravity}},\ }\href
  {https://doi.org/10.1007/JHEP06(2011)135} {\bibfield  {journal} {\bibinfo
  {journal} {JHEP}\ }\textbf {\bibinfo {volume} {06}},\ \bibinfo {pages}
  {135}},\ \Eprint {https://arxiv.org/abs/1104.3188} {arXiv:1104.3188 [hep-th]}
  \BibitemShut {NoStop}%
\bibitem [{\citenamefont {Bazeia}\ \emph {et~al.}(2014)\citenamefont {Bazeia},
  \citenamefont {Lob{\~a}o}, \citenamefont {Menezes}, \citenamefont {Petrov},\
  and\ \citenamefont {da~Silva}}]{Bazeia:2013uva}%
  \BibitemOpen
  \bibfield  {author} {\bibinfo {author} {\bibfnamefont {D.}~\bibnamefont
  {Bazeia}}, \bibinfo {author} {\bibfnamefont {A.~S.}\ \bibnamefont
  {Lob{\~a}o}, \bibfnamefont {Jr.}}, \bibinfo {author} {\bibfnamefont
  {R.}~\bibnamefont {Menezes}}, \bibinfo {author} {\bibfnamefont {A.~Y.}\
  \bibnamefont {Petrov}},\ and\ \bibinfo {author} {\bibfnamefont {A.~J.}\
  \bibnamefont {da~Silva}},\ }\bibfield  {title} {\bibinfo {title} {{Braneworld
  solutions for $f(R)$ models with non-constant curvature}},\ }\href
  {https://doi.org/10.1016/j.physletb.2014.01.011} {\bibfield  {journal}
  {\bibinfo  {journal} {Phys. Lett. B}\ }\textbf {\bibinfo {volume} {729}},\
  \bibinfo {pages} {127} (\bibinfo {year} {2014})},\ \Eprint
  {https://arxiv.org/abs/1311.6294} {arXiv:1311.6294 [hep-th]} \BibitemShut
  {NoStop}%
\bibitem [{\citenamefont {Geng}\ and\ \citenamefont {Lu}(2016)}]{Geng}%
  \BibitemOpen
  \bibfield  {author} {\bibinfo {author} {\bibfnamefont {W.-J.}\ \bibnamefont
  {Geng}}\ and\ \bibinfo {author} {\bibfnamefont {H.}~\bibnamefont {Lu}},\
  }\bibfield  {title} {\bibinfo {title} {{Einstein-Vector Gravity, Emerging
  Gauge Symmetry and de Sitter Bounce}},\ }\href
  {https://doi.org/10.1103/PhysRevD.93.044035} {\bibfield  {journal} {\bibinfo
  {journal} {Phys. Rev. D}\ }\textbf {\bibinfo {volume} {93}},\ \bibinfo
  {pages} {044035} (\bibinfo {year} {2016})},\ \Eprint
  {https://arxiv.org/abs/1511.03681} {arXiv:1511.03681 [hep-th]} \BibitemShut
  {NoStop}%
\bibitem [{\citenamefont {Gu}\ \emph {et~al.}(2017)\citenamefont {Gu},
  \citenamefont {Zhang}, \citenamefont {Yu},\ and\ \citenamefont
  {Liu}}]{Gu:2016nyo}%
  \BibitemOpen
  \bibfield  {author} {\bibinfo {author} {\bibfnamefont {B.-M.}\ \bibnamefont
  {Gu}}, \bibinfo {author} {\bibfnamefont {Y.-P.}\ \bibnamefont {Zhang}},
  \bibinfo {author} {\bibfnamefont {H.}~\bibnamefont {Yu}},\ and\ \bibinfo
  {author} {\bibfnamefont {Y.-X.}\ \bibnamefont {Liu}},\ }\bibfield  {title}
  {\bibinfo {title} {{Full linear perturbations and localization of gravity on
  $f(R,T)$ brane}},\ }\href {https://doi.org/10.1140/epjc/s10052-017-4666-3}
  {\bibfield  {journal} {\bibinfo  {journal} {Eur. Phys. J. C}\ }\textbf
  {\bibinfo {volume} {77}},\ \bibinfo {pages} {115} (\bibinfo {year} {2017})},\
  \Eprint {https://arxiv.org/abs/1606.07169} {arXiv:1606.07169 [hep-th]}
  \BibitemShut {NoStop}%
\bibitem [{\citenamefont {Zhong}\ and\ \citenamefont
  {Liu}(2017)}]{Zhong:2016iko}%
  \BibitemOpen
  \bibfield  {author} {\bibinfo {author} {\bibfnamefont {Y.}~\bibnamefont
  {Zhong}}\ and\ \bibinfo {author} {\bibfnamefont {Y.-X.}\ \bibnamefont
  {Liu}},\ }\bibfield  {title} {\bibinfo {title} {{Linearization of a warped
  $f(R)$ theory in the higher-order frame}},\ }\href
  {https://doi.org/10.1103/PhysRevD.95.104060} {\bibfield  {journal} {\bibinfo
  {journal} {Phys. Rev. D}\ }\textbf {\bibinfo {volume} {95}},\ \bibinfo
  {pages} {104060} (\bibinfo {year} {2017})},\ \Eprint
  {https://arxiv.org/abs/1611.08237} {arXiv:1611.08237 [gr-qc]} \BibitemShut
  {NoStop}%
\bibitem [{\citenamefont {Zhong}\ \emph {et~al.}(2018)\citenamefont {Zhong},
  \citenamefont {Yang},\ and\ \citenamefont {Liu}}]{Zhong:2017ffr}%
  \BibitemOpen
  \bibfield  {author} {\bibinfo {author} {\bibfnamefont {Y.}~\bibnamefont
  {Zhong}}, \bibinfo {author} {\bibfnamefont {K.}~\bibnamefont {Yang}},\ and\
  \bibinfo {author} {\bibfnamefont {Y.-X.}\ \bibnamefont {Liu}},\ }\bibfield
  {title} {\bibinfo {title} {{Linearization of a warped $f(R)$ theory in the
  higher-order frame II: the equation of motion approach}},\ }\href
  {https://doi.org/10.1103/PhysRevD.97.044032} {\bibfield  {journal} {\bibinfo
  {journal} {Phys. Rev. D}\ }\textbf {\bibinfo {volume} {97}},\ \bibinfo
  {pages} {044032} (\bibinfo {year} {2018})},\ \Eprint
  {https://arxiv.org/abs/1708.03737} {arXiv:1708.03737 [gr-qc]} \BibitemShut
  {NoStop}%
\bibitem [{\citenamefont {Zhou}\ \emph {et~al.}(2018)\citenamefont {Zhou},
  \citenamefont {Du}, \citenamefont {Yu},\ and\ \citenamefont
  {Liu}}]{Zhou:2017xaq}%
  \BibitemOpen
  \bibfield  {author} {\bibinfo {author} {\bibfnamefont {X.-N.}\ \bibnamefont
  {Zhou}}, \bibinfo {author} {\bibfnamefont {Y.-Z.}\ \bibnamefont {Du}},
  \bibinfo {author} {\bibfnamefont {H.}~\bibnamefont {Yu}},\ and\ \bibinfo
  {author} {\bibfnamefont {Y.-X.}\ \bibnamefont {Liu}},\ }\bibfield  {title}
  {\bibinfo {title} {{Localization of Gravitino Field on $f(R)$ Thick
  Branes}},\ }\href {https://doi.org/10.1007/s11433-018-9246-2} {\bibfield
  {journal} {\bibinfo  {journal} {Sci. China Phys. Mech. Astron.}\ }\textbf
  {\bibinfo {volume} {61}},\ \bibinfo {pages} {110411} (\bibinfo {year}
  {2018})},\ \Eprint {https://arxiv.org/abs/1703.10805} {arXiv:1703.10805
  [hep-th]} \BibitemShut {NoStop}%
\bibitem [{\citenamefont {Xie}\ \emph {et~al.}(2021)\citenamefont {Xie},
  \citenamefont {Fu}, \citenamefont {Sui}, \citenamefont {Zhao},\ and\
  \citenamefont {Zhong}}]{Xie:2021ayr}%
  \BibitemOpen
  \bibfield  {author} {\bibinfo {author} {\bibfnamefont {Q.-Y.}\ \bibnamefont
  {Xie}}, \bibinfo {author} {\bibfnamefont {Q.-M.}\ \bibnamefont {Fu}},
  \bibinfo {author} {\bibfnamefont {T.-T.}\ \bibnamefont {Sui}}, \bibinfo
  {author} {\bibfnamefont {L.}~\bibnamefont {Zhao}},\ and\ \bibinfo {author}
  {\bibfnamefont {Y.}~\bibnamefont {Zhong}},\ }\bibfield  {title} {\bibinfo
  {title} {{First-Order Formalism and Thick Branes in Mimetic Gravity}},\
  }\href {https://doi.org/10.3390/sym13081345} {\bibfield  {journal} {\bibinfo
  {journal} {Symmetry}\ }\textbf {\bibinfo {volume} {13}},\ \bibinfo {pages}
  {1345} (\bibinfo {year} {2021})},\ \Eprint {https://arxiv.org/abs/2102.10251}
  {arXiv:2102.10251 [gr-qc]} \BibitemShut {NoStop}%
\bibitem [{\citenamefont {Moreira}\ \emph {et~al.}(2021)\citenamefont
  {Moreira}, \citenamefont {Lima}, \citenamefont {Silva},\ and\ \citenamefont
  {Almeida}}]{Moreira:2021uod}%
  \BibitemOpen
  \bibfield  {author} {\bibinfo {author} {\bibfnamefont {A.~R.~P.}\
  \bibnamefont {Moreira}}, \bibinfo {author} {\bibfnamefont {F.~C.~E.}\
  \bibnamefont {Lima}}, \bibinfo {author} {\bibfnamefont {J.~E.~G.}\
  \bibnamefont {Silva}},\ and\ \bibinfo {author} {\bibfnamefont {C.~A.~S.}\
  \bibnamefont {Almeida}},\ }\bibfield  {title} {\bibinfo {title} {{First-order
  formalism for thick branes in $f(T,{\mathscr {T}})$ gravity}},\ }\href
  {https://doi.org/10.1140/epjc/s10052-021-09883-2} {\bibfield  {journal}
  {\bibinfo  {journal} {Eur. Phys. J. C}\ }\textbf {\bibinfo {volume} {81}},\
  \bibinfo {pages} {1081} (\bibinfo {year} {2021})},\ \Eprint
  {https://arxiv.org/abs/2107.04142} {arXiv:2107.04142 [gr-qc]} \BibitemShut
  {NoStop}%
\bibitem [{\citenamefont {Xu}\ \emph {et~al.}(2023)\citenamefont {Xu},
  \citenamefont {Chen}, \citenamefont {Zhang},\ and\ \citenamefont
  {Liu}}]{Xu:2022xxd}%
  \BibitemOpen
  \bibfield  {author} {\bibinfo {author} {\bibfnamefont {N.}~\bibnamefont
  {Xu}}, \bibinfo {author} {\bibfnamefont {J.}~\bibnamefont {Chen}}, \bibinfo
  {author} {\bibfnamefont {Y.-P.}\ \bibnamefont {Zhang}},\ and\ \bibinfo
  {author} {\bibfnamefont {Y.-X.}\ \bibnamefont {Liu}},\ }\bibfield  {title}
  {\bibinfo {title} {{Multikink brane in Gauss-Bonnet gravity and its
  stability}},\ }\href {https://doi.org/10.1103/PhysRevD.107.124011} {\bibfield
   {journal} {\bibinfo  {journal} {Phys. Rev. D}\ }\textbf {\bibinfo {volume}
  {107}},\ \bibinfo {pages} {124011} (\bibinfo {year} {2023})},\ \Eprint
  {https://arxiv.org/abs/2201.10282} {arXiv:2201.10282 [hep-th]} \BibitemShut
  {NoStop}%
\bibitem [{\citenamefont {Silva}\ \emph {et~al.}(2022)\citenamefont {Silva},
  \citenamefont {Maluf}, \citenamefont {Olmo},\ and\ \citenamefont
  {Almeida}}]{Silva:2022pfd}%
  \BibitemOpen
  \bibfield  {author} {\bibinfo {author} {\bibfnamefont {J.~E.~G.}\
  \bibnamefont {Silva}}, \bibinfo {author} {\bibfnamefont {R.~V.}\ \bibnamefont
  {Maluf}}, \bibinfo {author} {\bibfnamefont {G.~J.}\ \bibnamefont {Olmo}},\
  and\ \bibinfo {author} {\bibfnamefont {C.~A.~S.}\ \bibnamefont {Almeida}},\
  }\bibfield  {title} {\bibinfo {title} {{Braneworlds in $f(Q)$ gravity}},\
  }\href {https://doi.org/10.1103/PhysRevD.106.024033} {\bibfield  {journal}
  {\bibinfo  {journal} {Phys. Rev. D}\ }\textbf {\bibinfo {volume} {106}},\
  \bibinfo {pages} {024033} (\bibinfo {year} {2022})},\ \Eprint
  {https://arxiv.org/abs/2203.05720} {arXiv:2203.05720 [gr-qc]} \BibitemShut
  {NoStop}%
\bibitem [{\citenamefont {Xu}\ and\ \citenamefont {Zhang}(2022)}]{Xu:2022gth}%
  \BibitemOpen
  \bibfield  {author} {\bibinfo {author} {\bibfnamefont {Y.}~\bibnamefont
  {Xu}}\ and\ \bibinfo {author} {\bibfnamefont {X.}~\bibnamefont {Zhang}},\
  }\bibfield  {title} {\bibinfo {title} {{Tensor Perturbations and Thick Branes
  in Higher Dimensional Gauss-Bonnet Gravity}},\ }\href@noop {} {\  (\bibinfo
  {year} {2022})},\ \Eprint {https://arxiv.org/abs/2203.13401}
  {arXiv:2203.13401 [hep-th]} \BibitemShut {NoStop}%
\bibitem [{\citenamefont {Dzhunushaliev}\ \emph {et~al.}(2010)\citenamefont
  {Dzhunushaliev}, \citenamefont {Folomeev},\ and\ \citenamefont
  {Minamitsuji}}]{Dzhunushaliev:2009va}%
  \BibitemOpen
  \bibfield  {author} {\bibinfo {author} {\bibfnamefont {V.}~\bibnamefont
  {Dzhunushaliev}}, \bibinfo {author} {\bibfnamefont {V.}~\bibnamefont
  {Folomeev}},\ and\ \bibinfo {author} {\bibfnamefont {M.}~\bibnamefont
  {Minamitsuji}},\ }\bibfield  {title} {\bibinfo {title} {{Thick brane
  solutions}},\ }\href {https://doi.org/10.1088/0034-4885/73/6/066901}
  {\bibfield  {journal} {\bibinfo  {journal} {Rept. Prog. Phys.}\ }\textbf
  {\bibinfo {volume} {73}},\ \bibinfo {pages} {066901} (\bibinfo {year}
  {2010})},\ \Eprint {https://arxiv.org/abs/0904.1775} {arXiv:0904.1775
  [gr-qc]} \BibitemShut {NoStop}%
\bibitem [{\citenamefont {Maartens}\ and\ \citenamefont
  {Koyama}(2010)}]{Maartens:2010ar}%
  \BibitemOpen
  \bibfield  {author} {\bibinfo {author} {\bibfnamefont {R.}~\bibnamefont
  {Maartens}}\ and\ \bibinfo {author} {\bibfnamefont {K.}~\bibnamefont
  {Koyama}},\ }\bibfield  {title} {\bibinfo {title} {{Brane-World Gravity}},\
  }\href {https://doi.org/10.12942/lrr-2010-5} {\bibfield  {journal} {\bibinfo
  {journal} {Living Rev. Rel.}\ }\textbf {\bibinfo {volume} {13}},\ \bibinfo
  {pages} {5} (\bibinfo {year} {2010})},\ \Eprint
  {https://arxiv.org/abs/1004.3962} {arXiv:1004.3962 [hep-th]} \BibitemShut
  {NoStop}%
\bibitem [{\citenamefont {Liu}(2018)}]{Liu:2017gcn}%
  \BibitemOpen
  \bibfield  {author} {\bibinfo {author} {\bibfnamefont {Y.-X.}\ \bibnamefont
  {Liu}},\ }\bibinfo {title} {{Introduction to Extra Dimensions and Thick
  Braneworlds},}\ (\bibinfo {year} {2018})\ \Eprint
  {https://arxiv.org/abs/1707.08541} {arXiv:1707.08541 [hep-th]} \BibitemShut
  {NoStop}%
\bibitem [{\citenamefont {Ahluwalia}\ \emph {et~al.}(2022)\citenamefont
  {Ahluwalia}, \citenamefont {da~Silva}, \citenamefont {Lee}, \citenamefont
  {Liu}, \citenamefont {Pereira},\ and\ \citenamefont
  {Sorkhi}}]{Ahluwalia:2022ttu}%
  \BibitemOpen
  \bibfield  {author} {\bibinfo {author} {\bibfnamefont {D.~V.}\ \bibnamefont
  {Ahluwalia}}, \bibinfo {author} {\bibfnamefont {J.~M.~H.}\ \bibnamefont
  {da~Silva}}, \bibinfo {author} {\bibfnamefont {C.-Y.}\ \bibnamefont {Lee}},
  \bibinfo {author} {\bibfnamefont {Y.-X.}\ \bibnamefont {Liu}}, \bibinfo
  {author} {\bibfnamefont {S.~H.}\ \bibnamefont {Pereira}},\ and\ \bibinfo
  {author} {\bibfnamefont {M.~M.}\ \bibnamefont {Sorkhi}},\ }\bibfield  {title}
  {\bibinfo {title} {{Mass dimension one fermions: Constructing darkness}},\
  }\href {https://doi.org/10.1016/j.physrep.2022.04.003} {\bibfield  {journal}
  {\bibinfo  {journal} {Phys. Rept.}\ }\textbf {\bibinfo {volume} {967}},\
  \bibinfo {pages} {1} (\bibinfo {year} {2022})},\ \Eprint
  {https://arxiv.org/abs/2205.04754} {arXiv:2205.04754 [hep-ph]} \BibitemShut
  {NoStop}%
\bibitem [{\citenamefont {Kokkotas}\ and\ \citenamefont
  {Schmidt}(1999)}]{Kokkotas:1999bd}%
  \BibitemOpen
  \bibfield  {author} {\bibinfo {author} {\bibfnamefont {K.~D.}\ \bibnamefont
  {Kokkotas}}\ and\ \bibinfo {author} {\bibfnamefont {B.~G.}\ \bibnamefont
  {Schmidt}},\ }\bibfield  {title} {\bibinfo {title} {{Quasinormal modes of
  stars and black holes}},\ }\href {https://doi.org/10.12942/lrr-1999-2}
  {\bibfield  {journal} {\bibinfo  {journal} {Living Rev. Rel.}\ }\textbf
  {\bibinfo {volume} {2}},\ \bibinfo {pages} {2} (\bibinfo {year} {1999})},\
  \Eprint {https://arxiv.org/abs/gr-qc/9909058} {arXiv:gr-qc/9909058}
  \BibitemShut {NoStop}%
\bibitem [{\citenamefont {Nollert}(1999)}]{Nollert:1999ji}%
  \BibitemOpen
  \bibfield  {author} {\bibinfo {author} {\bibfnamefont {H.-P.}\ \bibnamefont
  {Nollert}},\ }\bibfield  {title} {\bibinfo {title} {{Quasinormal modes: the
  characteristic ``sound'' of black holes and neutron stars}},\ }\href
  {https://doi.org/10.1088/0264-9381/16/12/201} {\bibfield  {journal} {\bibinfo
   {journal} {Class. Quantum Grav.}\ }\textbf {\bibinfo {volume} {16}},\
  \bibinfo {pages} {R159} (\bibinfo {year} {1999})}\BibitemShut {NoStop}%
\bibitem [{\citenamefont {Berti}\ \emph {et~al.}(2009)\citenamefont {Berti},
  \citenamefont {Cardoso},\ and\ \citenamefont {Starinets}}]{Berti:2009kk}%
  \BibitemOpen
  \bibfield  {author} {\bibinfo {author} {\bibfnamefont {E.}~\bibnamefont
  {Berti}}, \bibinfo {author} {\bibfnamefont {V.}~\bibnamefont {Cardoso}},\
  and\ \bibinfo {author} {\bibfnamefont {A.~O.}\ \bibnamefont {Starinets}},\
  }\bibfield  {title} {\bibinfo {title} {{Quasinormal modes of black holes and
  black branes}},\ }\href {https://doi.org/10.1088/0264-9381/26/16/163001}
  {\bibfield  {journal} {\bibinfo  {journal} {Class. Quant. Grav.}\ }\textbf
  {\bibinfo {volume} {26}},\ \bibinfo {pages} {163001} (\bibinfo {year}
  {2009})},\ \Eprint {https://arxiv.org/abs/0905.2975} {arXiv:0905.2975
  [gr-qc]} \BibitemShut {NoStop}%
\bibitem [{\citenamefont {Konoplya}\ and\ \citenamefont
  {Zhidenko}(2011)}]{Konoplya:2011qq}%
  \BibitemOpen
  \bibfield  {author} {\bibinfo {author} {\bibfnamefont {R.~A.}\ \bibnamefont
  {Konoplya}}\ and\ \bibinfo {author} {\bibfnamefont {A.}~\bibnamefont
  {Zhidenko}},\ }\bibfield  {title} {\bibinfo {title} {{Quasinormal modes of
  black holes: From astrophysics to string theory}},\ }\href
  {https://doi.org/10.1103/RevModPhys.83.793} {\bibfield  {journal} {\bibinfo
  {journal} {Rev. Mod. Phys.}\ }\textbf {\bibinfo {volume} {83}},\ \bibinfo
  {pages} {793} (\bibinfo {year} {2011})},\ \Eprint
  {https://arxiv.org/abs/1102.4014} {arXiv:1102.4014 [gr-qc]} \BibitemShut
  {NoStop}%
\bibitem [{\citenamefont {Cardoso}\ \emph {et~al.}(2016)\citenamefont
  {Cardoso}, \citenamefont {ranzin},\ and\ \citenamefont
  {Pani}}]{Cardoso:2016rao}%
  \BibitemOpen
  \bibfield  {author} {\bibinfo {author} {\bibfnamefont {V.}~\bibnamefont
  {Cardoso}}, \bibinfo {author} {\bibfnamefont {E.}~\bibnamefont {ranzin}},\
  and\ \bibinfo {author} {\bibfnamefont {P.}~\bibnamefont {Pani}},\ }\bibfield
  {title} {\bibinfo {title} {{Is the gravitational-wave ringdown a probe of the
  event horizon?}},\ }\href {https://doi.org/10.1103/PhysRevLett.116.171101}
  {\bibfield  {journal} {\bibinfo  {journal} {Phys. Rev. Lett.}\ }\textbf
  {\bibinfo {volume} {116}},\ \bibinfo {pages} {171101} (\bibinfo {year}
  {2016})},\ \bibinfo {note} {[Erratum: Phys.Rev.Lett. 117, 089902 (2016)]},\
  \Eprint {https://arxiv.org/abs/1602.07309} {arXiv:1602.07309 [gr-qc]}
  \BibitemShut {NoStop}%
\bibitem [{\citenamefont {Jusufi}\ \emph {et~al.}(2021)\citenamefont {Jusufi},
  \citenamefont {Azreg-A{\"\i}nou}, \citenamefont {Jamil}, \citenamefont {Wei},
  \citenamefont {Wu},\ and\ \citenamefont {Wang}}]{Jusufi:2020odz}%
  \BibitemOpen
  \bibfield  {author} {\bibinfo {author} {\bibfnamefont {K.}~\bibnamefont
  {Jusufi}}, \bibinfo {author} {\bibfnamefont {M.}~\bibnamefont
  {Azreg-A{\"\i}nou}}, \bibinfo {author} {\bibfnamefont {M.}~\bibnamefont
  {Jamil}}, \bibinfo {author} {\bibfnamefont {S.-W.}\ \bibnamefont {Wei}},
  \bibinfo {author} {\bibfnamefont {Q.}~\bibnamefont {Wu}},\ and\ \bibinfo
  {author} {\bibfnamefont {A.}~\bibnamefont {Wang}},\ }\bibfield  {title}
  {\bibinfo {title} {{Quasinormal modes, quasiperiodic oscillations, and the
  shadow of rotating regular black holes in nonminimally coupled
  Einstein-Yang-Mills theory}},\ }\href
  {https://doi.org/10.1103/PhysRevD.103.024013} {\bibfield  {journal} {\bibinfo
   {journal} {Phys. Rev. D}\ }\textbf {\bibinfo {volume} {103}},\ \bibinfo
  {pages} {024013} (\bibinfo {year} {2021})},\ \Eprint
  {https://arxiv.org/abs/2008.08450} {arXiv:2008.08450 [gr-qc]} \BibitemShut
  {NoStop}%
\bibitem [{\citenamefont {Cheung}\ \emph {et~al.}(2022)\citenamefont {Cheung},
  \citenamefont {Destounis}, \citenamefont {Macedo}, \citenamefont {Berti},\
  and\ \citenamefont {Cardoso}}]{Cheung:2021bol}%
  \BibitemOpen
  \bibfield  {author} {\bibinfo {author} {\bibfnamefont {M.~H.-Y.}\
  \bibnamefont {Cheung}}, \bibinfo {author} {\bibfnamefont {K.}~\bibnamefont
  {Destounis}}, \bibinfo {author} {\bibfnamefont {R.~P.}\ \bibnamefont
  {Macedo}}, \bibinfo {author} {\bibfnamefont {E.}~\bibnamefont {Berti}},\ and\
  \bibinfo {author} {\bibfnamefont {V.}~\bibnamefont {Cardoso}},\ }\bibfield
  {title} {\bibinfo {title} {{Destabilizing the Fundamental Mode of Black
  Holes: The Elephant and the Flea}},\ }\href
  {https://doi.org/10.1103/PhysRevLett.128.111103} {\bibfield  {journal}
  {\bibinfo  {journal} {Phys. Rev. Lett.}\ }\textbf {\bibinfo {volume} {128}},\
  \bibinfo {pages} {111103} (\bibinfo {year} {2022})},\ \Eprint
  {https://arxiv.org/abs/2111.05415} {arXiv:2111.05415 [gr-qc]} \BibitemShut
  {NoStop}%
\bibitem [{\citenamefont {Seahra}(2005{\natexlab{a}})}]{Seahra:2005iq}%
  \BibitemOpen
  \bibfield  {author} {\bibinfo {author} {\bibfnamefont {S.~S.}\ \bibnamefont
  {Seahra}},\ }\bibfield  {title} {\bibinfo {title} {{Metastable massive
  gravitons from an infinite extra dimension}},\ }\href
  {https://doi.org/10.1142/S0218271805007887} {\bibfield  {journal} {\bibinfo
  {journal} {Int. J. Mod. Phys. D}\ }\textbf {\bibinfo {volume} {14}},\
  \bibinfo {pages} {2279} (\bibinfo {year} {2005}{\natexlab{a}})},\ \Eprint
  {https://arxiv.org/abs/hep-th/0505196} {arXiv:hep-th/0505196} \BibitemShut
  {NoStop}%
\bibitem [{\citenamefont {Seahra}(2005{\natexlab{b}})}]{Seahra:2005wk}%
  \BibitemOpen
  \bibfield  {author} {\bibinfo {author} {\bibfnamefont {S.~S.}\ \bibnamefont
  {Seahra}},\ }\bibfield  {title} {\bibinfo {title} {{Ringing the
  Randall-Sundrum braneworld:Metastable gravity wave bound states}},\ }\href
  {https://doi.org/10.1103/PhysRevD.72.066002} {\bibfield  {journal} {\bibinfo
  {journal} {Phys. Rev. D}\ }\textbf {\bibinfo {volume} {72}},\ \bibinfo
  {pages} {066002} (\bibinfo {year} {2005}{\natexlab{b}})},\ \Eprint
  {https://arxiv.org/abs/hep-th/0501175} {arXiv:hep-th/0501175} \BibitemShut
  {NoStop}%
\bibitem [{\citenamefont {Deng}\ \emph {et~al.}(2025)\citenamefont {Deng},
  \citenamefont {Long}, \citenamefont {Tan}, \citenamefont {Chen},\ and\
  \citenamefont {Jing}}]{Deng1}%
  \BibitemOpen
  \bibfield  {author} {\bibinfo {author} {\bibfnamefont {W.}~\bibnamefont
  {Deng}}, \bibinfo {author} {\bibfnamefont {S.}~\bibnamefont {Long}}, \bibinfo
  {author} {\bibfnamefont {Q.}~\bibnamefont {Tan}}, \bibinfo {author}
  {\bibfnamefont {Z.-C.}\ \bibnamefont {Chen}},\ and\ \bibinfo {author}
  {\bibfnamefont {J.}~\bibnamefont {Jing}},\ }\bibfield  {title} {\bibinfo
  {title} {{Scalar-gravitational quasinormal modes and echoes in a five
  dimensional thick brane}},\ }\href@noop {} {\  (\bibinfo {year} {2025})},\
  \Eprint {https://arxiv.org/abs/2508.20937} {arXiv:2508.20937 [gr-qc]}
  \BibitemShut {NoStop}%
\bibitem [{\citenamefont {Yang}\ \emph {et~al.}(2024)\citenamefont {Yang},
  \citenamefont {Guo}, \citenamefont {Tan}, \citenamefont {Zhao},\ and\
  \citenamefont {Liu}}]{Yang1}%
  \BibitemOpen
  \bibfield  {author} {\bibinfo {author} {\bibfnamefont {S.}~\bibnamefont
  {Yang}}, \bibinfo {author} {\bibfnamefont {W.-D.}\ \bibnamefont {Guo}},
  \bibinfo {author} {\bibfnamefont {Q.}~\bibnamefont {Tan}}, \bibinfo {author}
  {\bibfnamefont {L.}~\bibnamefont {Zhao}},\ and\ \bibinfo {author}
  {\bibfnamefont {Y.-X.}\ \bibnamefont {Liu}},\ }\bibfield  {title} {\bibinfo
  {title} {{Parametrized quasinormal frequencies and Hawking radiation for
  axial gravitational perturbations of a holonomy-corrected black hole}},\
  }\href {https://doi.org/10.1103/PhysRevD.110.064051} {\bibfield  {journal}
  {\bibinfo  {journal} {Phys. Rev. D}\ }\textbf {\bibinfo {volume} {110}},\
  \bibinfo {pages} {064051} (\bibinfo {year} {2024})},\ \Eprint
  {https://arxiv.org/abs/2406.15711} {arXiv:2406.15711 [gr-qc]} \BibitemShut
  {NoStop}%
\bibitem [{\citenamefont {Jia}\ \emph {et~al.}(2025)\citenamefont {Jia},
  \citenamefont {Guo}, \citenamefont {Liu},\ and\ \citenamefont {Tan}}]{Jia1}%
  \BibitemOpen
  \bibfield  {author} {\bibinfo {author} {\bibfnamefont {H.-L.}\ \bibnamefont
  {Jia}}, \bibinfo {author} {\bibfnamefont {W.-D.}\ \bibnamefont {Guo}},
  \bibinfo {author} {\bibfnamefont {Y.-X.}\ \bibnamefont {Liu}},\ and\ \bibinfo
  {author} {\bibfnamefont {Q.}~\bibnamefont {Tan}},\ }\bibfield  {title}
  {\bibinfo {title} {{Quasinormal ringing of de Sitter braneworlds}},\ }\href
  {https://doi.org/10.1007/JHEP06(2025)117} {\bibfield  {journal} {\bibinfo
  {journal} {JHEP}\ }\textbf {\bibinfo {volume} {06}},\ \bibinfo {pages}
  {117}},\ \Eprint {https://arxiv.org/abs/2501.00477} {arXiv:2501.00477
  [gr-qc]} \BibitemShut {NoStop}%
\bibitem [{\citenamefont {Jia}\ \emph {et~al.}(2024)\citenamefont {Jia},
  \citenamefont {Guo}, \citenamefont {Tan},\ and\ \citenamefont {Liu}}]{Jia2}%
  \BibitemOpen
  \bibfield  {author} {\bibinfo {author} {\bibfnamefont {H.-L.}\ \bibnamefont
  {Jia}}, \bibinfo {author} {\bibfnamefont {W.-D.}\ \bibnamefont {Guo}},
  \bibinfo {author} {\bibfnamefont {Q.}~\bibnamefont {Tan}},\ and\ \bibinfo
  {author} {\bibfnamefont {Y.-X.}\ \bibnamefont {Liu}},\ }\bibfield  {title}
  {\bibinfo {title} {{Quasinormal ringing of thick braneworlds with a finite
  extra dimension}},\ }\href {https://doi.org/10.1103/PhysRevD.110.064077}
  {\bibfield  {journal} {\bibinfo  {journal} {Phys. Rev. D}\ }\textbf {\bibinfo
  {volume} {110}},\ \bibinfo {pages} {064077} (\bibinfo {year} {2024})},\
  \Eprint {https://arxiv.org/abs/2406.03929} {arXiv:2406.03929 [gr-qc]}
  \BibitemShut {NoStop}%
\bibitem [{\citenamefont {Guo}\ \emph {et~al.}(2024)\citenamefont {Guo},
  \citenamefont {Tan},\ and\ \citenamefont {Liu}}]{Guo1}%
  \BibitemOpen
  \bibfield  {author} {\bibinfo {author} {\bibfnamefont {W.-D.}\ \bibnamefont
  {Guo}}, \bibinfo {author} {\bibfnamefont {Q.}~\bibnamefont {Tan}},\ and\
  \bibinfo {author} {\bibfnamefont {Y.-X.}\ \bibnamefont {Liu}},\ }\bibfield
  {title} {\bibinfo {title} {{Quasinormal modes and greybody factor of a
  Lorentz-violating black hole}},\ }\href
  {https://doi.org/10.1088/1475-7516/2024/07/008} {\bibfield  {journal}
  {\bibinfo  {journal} {JCAP}\ }\textbf {\bibinfo {volume} {07}},\ \bibinfo
  {pages} {008}},\ \Eprint {https://arxiv.org/abs/2312.16605} {arXiv:2312.16605
  [gr-qc]} \BibitemShut {NoStop}%
\bibitem [{\citenamefont {Guo}\ and\ \citenamefont {Tan}(2023)}]{Guo2}%
  \BibitemOpen
  \bibfield  {author} {\bibinfo {author} {\bibfnamefont {W.-D.}\ \bibnamefont
  {Guo}}\ and\ \bibinfo {author} {\bibfnamefont {Q.}~\bibnamefont {Tan}},\
  }\bibfield  {title} {\bibinfo {title} {{Quasinormal Modes of a Charged Black
  Hole with Scalar Hair}},\ }\href {https://doi.org/10.3390/universe9070320}
  {\bibfield  {journal} {\bibinfo  {journal} {Universe}\ }\textbf {\bibinfo
  {volume} {9}},\ \bibinfo {pages} {320} (\bibinfo {year} {2023})},\ \Eprint
  {https://arxiv.org/abs/2402.14265} {arXiv:2402.14265 [gr-qc]} \BibitemShut
  {NoStop}%
\bibitem [{\citenamefont {Tan}\ \emph {et~al.}(2023)\citenamefont {Tan},
  \citenamefont {Zhang}, \citenamefont {Guo}, \citenamefont {Chen},
  \citenamefont {Zhu},\ and\ \citenamefont {Liu}}]{Tan:2022uex}%
  \BibitemOpen
  \bibfield  {author} {\bibinfo {author} {\bibfnamefont {Q.}~\bibnamefont
  {Tan}}, \bibinfo {author} {\bibfnamefont {Y.-P.}\ \bibnamefont {Zhang}},
  \bibinfo {author} {\bibfnamefont {W.-D.}\ \bibnamefont {Guo}}, \bibinfo
  {author} {\bibfnamefont {J.}~\bibnamefont {Chen}}, \bibinfo {author}
  {\bibfnamefont {C.-C.}\ \bibnamefont {Zhu}},\ and\ \bibinfo {author}
  {\bibfnamefont {Y.-X.}\ \bibnamefont {Liu}},\ }\bibfield  {title} {\bibinfo
  {title} {{Evolution of scalar field resonances in a braneworld}},\ }\href
  {https://doi.org/10.1140/epjc/s10052-022-11164-5} {\bibfield  {journal}
  {\bibinfo  {journal} {Eur. Phys. J. C}\ }\textbf {\bibinfo {volume} {83}},\
  \bibinfo {pages} {84} (\bibinfo {year} {2023})},\ \Eprint
  {https://arxiv.org/abs/2203.00277} {arXiv:2203.00277 [hep-th]} \BibitemShut
  {NoStop}%
\bibitem [{\citenamefont {Tan}\ \emph {et~al.}(2022)\citenamefont {Tan},
  \citenamefont {Guo},\ and\ \citenamefont {Liu}}]{Tan5}%
  \BibitemOpen
  \bibfield  {author} {\bibinfo {author} {\bibfnamefont {Q.}~\bibnamefont
  {Tan}}, \bibinfo {author} {\bibfnamefont {W.-D.}\ \bibnamefont {Guo}},\ and\
  \bibinfo {author} {\bibfnamefont {Y.-X.}\ \bibnamefont {Liu}},\ }\bibfield
  {title} {\bibinfo {title} {{Sound from extra dimensions: Quasinormal modes of
  a thick brane}},\ }\href {https://doi.org/10.1103/PhysRevD.106.044038}
  {\bibfield  {journal} {\bibinfo  {journal} {Phys. Rev. D}\ }\textbf {\bibinfo
  {volume} {106}},\ \bibinfo {pages} {044038} (\bibinfo {year} {2022})},\
  \Eprint {https://arxiv.org/abs/2205.05255} {arXiv:2205.05255 [gr-qc]}
  \BibitemShut {NoStop}%
\bibitem [{\citenamefont {Tan}\ \emph {et~al.}(2024{\natexlab{a}})\citenamefont
  {Tan}, \citenamefont {Guo}, \citenamefont {Zhang},\ and\ \citenamefont
  {Liu}}]{Tan4}%
  \BibitemOpen
  \bibfield  {author} {\bibinfo {author} {\bibfnamefont {Q.}~\bibnamefont
  {Tan}}, \bibinfo {author} {\bibfnamefont {W.-D.}\ \bibnamefont {Guo}},
  \bibinfo {author} {\bibfnamefont {Y.-P.}\ \bibnamefont {Zhang}},\ and\
  \bibinfo {author} {\bibfnamefont {Y.-X.}\ \bibnamefont {Liu}},\ }\bibfield
  {title} {\bibinfo {title} {{Characteristic modes of a thick brane: Resonances
  and quasinormal modes}},\ }\href
  {https://doi.org/10.1103/PhysRevD.109.024017} {\bibfield  {journal} {\bibinfo
   {journal} {Phys. Rev. D}\ }\textbf {\bibinfo {volume} {109}},\ \bibinfo
  {pages} {024017} (\bibinfo {year} {2024}{\natexlab{a}})},\ \Eprint
  {https://arxiv.org/abs/2304.09363} {arXiv:2304.09363 [gr-qc]} \BibitemShut
  {NoStop}%
\bibitem [{\citenamefont {Tan}\ \emph {et~al.}(2024{\natexlab{b}})\citenamefont
  {Tan}, \citenamefont {Zhong},\ and\ \citenamefont {Guo}}]{Tan3}%
  \BibitemOpen
  \bibfield  {author} {\bibinfo {author} {\bibfnamefont {Q.}~\bibnamefont
  {Tan}}, \bibinfo {author} {\bibfnamefont {Y.}~\bibnamefont {Zhong}},\ and\
  \bibinfo {author} {\bibfnamefont {W.-D.}\ \bibnamefont {Guo}},\ }\bibfield
  {title} {\bibinfo {title} {{Quasibound and quasinormal modes of a thick brane
  in Rastall gravity}},\ }\href {https://doi.org/10.1007/JHEP07(2024)252}
  {\bibfield  {journal} {\bibinfo  {journal} {JHEP}\ }\textbf {\bibinfo
  {volume} {07}},\ \bibinfo {pages} {252}},\ \Eprint
  {https://arxiv.org/abs/2404.11217} {arXiv:2404.11217 [gr-qc]} \BibitemShut
  {NoStop}%
\bibitem [{\citenamefont {Tan}\ \emph {et~al.}(2025{\natexlab{a}})\citenamefont
  {Tan}, \citenamefont {Long}, \citenamefont {Deng},\ and\ \citenamefont
  {Jing}}]{Tan2}%
  \BibitemOpen
  \bibfield  {author} {\bibinfo {author} {\bibfnamefont {Q.}~\bibnamefont
  {Tan}}, \bibinfo {author} {\bibfnamefont {S.}~\bibnamefont {Long}}, \bibinfo
  {author} {\bibfnamefont {W.}~\bibnamefont {Deng}},\ and\ \bibinfo {author}
  {\bibfnamefont {J.}~\bibnamefont {Jing}},\ }\bibfield  {title} {\bibinfo
  {title} {{Graviscalar quasinormal modes and asymptotic tails of a thick
  brane}},\ }\href {https://doi.org/10.1016/j.physletb.2025.139667} {\bibfield
  {journal} {\bibinfo  {journal} {Phys. Lett. B}\ }\textbf {\bibinfo {volume}
  {868}},\ \bibinfo {pages} {139667} (\bibinfo {year} {2025}{\natexlab{a}})},\
  \Eprint {https://arxiv.org/abs/2409.06947} {arXiv:2409.06947 [gr-qc]}
  \BibitemShut {NoStop}%
\bibitem [{\citenamefont {Tan}\ \emph {et~al.}(2025{\natexlab{b}})\citenamefont
  {Tan}, \citenamefont {Long}, \citenamefont {Deng},\ and\ \citenamefont
  {Jing}}]{Tan1}%
  \BibitemOpen
  \bibfield  {author} {\bibinfo {author} {\bibfnamefont {Q.}~\bibnamefont
  {Tan}}, \bibinfo {author} {\bibfnamefont {S.}~\bibnamefont {Long}}, \bibinfo
  {author} {\bibfnamefont {W.}~\bibnamefont {Deng}},\ and\ \bibinfo {author}
  {\bibfnamefont {J.}~\bibnamefont {Jing}},\ }\bibfield  {title} {\bibinfo
  {title} {{Quasinormal modes and echoes of a double braneworld}},\ }\href
  {https://doi.org/10.1007/JHEP02(2025)055} {\bibfield  {journal} {\bibinfo
  {journal} {JHEP}\ }\textbf {\bibinfo {volume} {02}},\ \bibinfo {pages}
  {055}},\ \Eprint {https://arxiv.org/abs/2410.06945} {arXiv:2410.06945
  [gr-qc]} \BibitemShut {NoStop}%
\bibitem [{\citenamefont {Zhu}\ \emph {et~al.}(2025)\citenamefont {Zhu},
  \citenamefont {Chen}, \citenamefont {Guo},\ and\ \citenamefont
  {Liu}}]{Zhu:2024gvl}%
  \BibitemOpen
  \bibfield  {author} {\bibinfo {author} {\bibfnamefont {C.-C.}\ \bibnamefont
  {Zhu}}, \bibinfo {author} {\bibfnamefont {J.}~\bibnamefont {Chen}}, \bibinfo
  {author} {\bibfnamefont {W.-D.}\ \bibnamefont {Guo}},\ and\ \bibinfo {author}
  {\bibfnamefont {Y.-X.}\ \bibnamefont {Liu}},\ }\bibfield  {title} {\bibinfo
  {title} {{Gravitational echoes from braneworlds}},\ }\href
  {https://doi.org/10.1007/JHEP01(2025)010} {\bibfield  {journal} {\bibinfo
  {journal} {JHEP}\ }\textbf {\bibinfo {volume} {01}},\ \bibinfo {pages}
  {010}},\ \Eprint {https://arxiv.org/abs/2406.16256} {arXiv:2406.16256
  [gr-qc]} \BibitemShut {NoStop}%
\bibitem [{\citenamefont {E}\ \emph {et~al.}(2025)\citenamefont {E},
  \citenamefont {Zhu},\ and\ \citenamefont {Liu}}]{E:2025kic}%
  \BibitemOpen
  \bibfield  {author} {\bibinfo {author} {\bibfnamefont {Y.-P.}\ \bibnamefont
  {E}}, \bibinfo {author} {\bibfnamefont {C.-C.}\ \bibnamefont {Zhu}},\ and\
  \bibinfo {author} {\bibfnamefont {Y.-X.}\ \bibnamefont {Liu}},\ }\bibfield
  {title} {\bibinfo {title} {{Quasinormal modes of thick branes in $f(R)$
  gravity}},\ }\href@noop {} {\  (\bibinfo {year} {2025})},\ \Eprint
  {https://arxiv.org/abs/2512.17208} {arXiv:2512.17208 [gr-qc]} \BibitemShut
  {NoStop}%
\bibitem [{\citenamefont {Hayashi}\ and\ \citenamefont
  {Shirafuji}(1979)}]{Hayashi:1979wj}%
  \BibitemOpen
  \bibfield  {author} {\bibinfo {author} {\bibfnamefont {K.}~\bibnamefont
  {Hayashi}}\ and\ \bibinfo {author} {\bibfnamefont {T.}~\bibnamefont
  {Shirafuji}},\ }\bibfield  {title} {\bibinfo {title} {{New general
  relativity}},\ }\href {https://doi.org/10.1103/PhysRevD.19.3524} {\bibfield
  {journal} {\bibinfo  {journal} {Phys. Rev. D}\ }\textbf {\bibinfo {volume}
  {19}},\ \bibinfo {pages} {3524} (\bibinfo {year} {1979})}\BibitemShut
  {NoStop}%
\bibitem [{\citenamefont {Aldrovandi}\ and\ \citenamefont
  {Pereira}(2013)}]{Aldrovandi:2013wha}%
  \BibitemOpen
  \bibfield  {author} {\bibinfo {author} {\bibfnamefont {R.}~\bibnamefont
  {Aldrovandi}}\ and\ \bibinfo {author} {\bibfnamefont {J.~G.}\ \bibnamefont
  {Pereira}},\ }\href {https://doi.org/10.1007/978-94-007-5143-9} {\emph
  {\bibinfo {title} {{Teleparallel Gravity}: {An Introduction}}}}\ (\bibinfo
  {publisher} {Springer},\ \bibinfo {year} {2013})\BibitemShut {NoStop}%
\bibitem [{\citenamefont {Ferraro}\ and\ \citenamefont
  {Fiorini}(2007)}]{Ferraro:2006jd}%
  \BibitemOpen
  \bibfield  {author} {\bibinfo {author} {\bibfnamefont {R.}~\bibnamefont
  {Ferraro}}\ and\ \bibinfo {author} {\bibfnamefont {F.}~\bibnamefont
  {Fiorini}},\ }\bibfield  {title} {\bibinfo {title} {{Modified teleparallel
  gravity: Inflation without inflaton}},\ }\href
  {https://doi.org/10.1103/PhysRevD.75.084031} {\bibfield  {journal} {\bibinfo
  {journal} {Phys. Rev. D}\ }\textbf {\bibinfo {volume} {75}},\ \bibinfo
  {pages} {084031} (\bibinfo {year} {2007})},\ \Eprint
  {https://arxiv.org/abs/gr-qc/0610067} {arXiv:gr-qc/0610067} \BibitemShut
  {NoStop}%
\bibitem [{\citenamefont {Bengochea}\ and\ \citenamefont
  {Ferraro}(2009)}]{Bengochea:2008gz}%
  \BibitemOpen
  \bibfield  {author} {\bibinfo {author} {\bibfnamefont {G.~R.}\ \bibnamefont
  {Bengochea}}\ and\ \bibinfo {author} {\bibfnamefont {R.}~\bibnamefont
  {Ferraro}},\ }\bibfield  {title} {\bibinfo {title} {{Dark torsion as the
  cosmic speed-up}},\ }\href {https://doi.org/10.1103/PhysRevD.79.124019}
  {\bibfield  {journal} {\bibinfo  {journal} {Phys. Rev. D}\ }\textbf {\bibinfo
  {volume} {79}},\ \bibinfo {pages} {124019} (\bibinfo {year} {2009})},\
  \Eprint {https://arxiv.org/abs/0812.1205} {arXiv:0812.1205 [astro-ph]}
  \BibitemShut {NoStop}%
\bibitem [{\citenamefont {Linder}(2010)}]{Linder:2010py}%
  \BibitemOpen
  \bibfield  {author} {\bibinfo {author} {\bibfnamefont {E.~V.}\ \bibnamefont
  {Linder}},\ }\bibfield  {title} {\bibinfo {title} {{Einstein's Other Gravity
  and the Acceleration of the Universe}},\ }\href
  {https://doi.org/10.1103/PhysRevD.81.127301} {\bibfield  {journal} {\bibinfo
  {journal} {Phys. Rev. D}\ }\textbf {\bibinfo {volume} {81}},\ \bibinfo
  {pages} {127301} (\bibinfo {year} {2010})},\ \bibinfo {note} {[Erratum:
  Phys.Rev.D 82, 109902 (2010)]},\ \Eprint {https://arxiv.org/abs/1005.3039}
  {arXiv:1005.3039 [astro-ph.CO]} \BibitemShut {NoStop}%
\bibitem [{\citenamefont {Karami}\ and\ \citenamefont
  {Abdolmaleki}(2013)}]{Karami:2010bys}%
  \BibitemOpen
  \bibfield  {author} {\bibinfo {author} {\bibfnamefont {K.}~\bibnamefont
  {Karami}}\ and\ \bibinfo {author} {\bibfnamefont {A.}~\bibnamefont
  {Abdolmaleki}},\ }\bibfield  {title} {\bibinfo {title} {{$f(T)$ modified
  teleparallel gravity models as an alternative for holographic and new
  agegraphic dark energy models}},\ }\href
  {https://doi.org/10.1088/1674-4527/13/7/001} {\bibfield  {journal} {\bibinfo
  {journal} {Res. Astron. Astrophys.}\ }\textbf {\bibinfo {volume} {13}},\
  \bibinfo {pages} {757} (\bibinfo {year} {2013})},\ \Eprint
  {https://arxiv.org/abs/1009.2459} {arXiv:1009.2459 [gr-qc]} \BibitemShut
  {NoStop}%
\bibitem [{\citenamefont {Bamba}\ \emph {et~al.}(2011)\citenamefont {Bamba},
  \citenamefont {Geng}, \citenamefont {Lee},\ and\ \citenamefont
  {Luo}}]{Bamba:2010wb}%
  \BibitemOpen
  \bibfield  {author} {\bibinfo {author} {\bibfnamefont {K.}~\bibnamefont
  {Bamba}}, \bibinfo {author} {\bibfnamefont {C.-Q.}\ \bibnamefont {Geng}},
  \bibinfo {author} {\bibfnamefont {C.-C.}\ \bibnamefont {Lee}},\ and\ \bibinfo
  {author} {\bibfnamefont {L.-W.}\ \bibnamefont {Luo}},\ }\bibfield  {title}
  {\bibinfo {title} {{Equation of state for dark energy in $f(T)$ gravity}},\
  }\href {https://doi.org/10.1088/1475-7516/2011/01/021} {\bibfield  {journal}
  {\bibinfo  {journal} {JCAP}\ }\textbf {\bibinfo {volume} {01}},\ \bibinfo
  {pages} {021}},\ \Eprint {https://arxiv.org/abs/1011.0508} {arXiv:1011.0508
  [astro-ph.CO]} \BibitemShut {NoStop}%
\bibitem [{\citenamefont {Cai}\ \emph {et~al.}(2016)\citenamefont {Cai},
  \citenamefont {Capozziello}, \citenamefont {De~Laurentis},\ and\
  \citenamefont {Saridakis}}]{Cai}%
  \BibitemOpen
  \bibfield  {author} {\bibinfo {author} {\bibfnamefont {Y.-F.}\ \bibnamefont
  {Cai}}, \bibinfo {author} {\bibfnamefont {S.}~\bibnamefont {Capozziello}},
  \bibinfo {author} {\bibfnamefont {M.}~\bibnamefont {De~Laurentis}},\ and\
  \bibinfo {author} {\bibfnamefont {E.~N.}\ \bibnamefont {Saridakis}},\
  }\bibfield  {title} {\bibinfo {title} {{$f(T)$ teleparallel gravity and
  cosmology}},\ }\href {https://doi.org/10.1088/0034-4885/79/10/106901}
  {\bibfield  {journal} {\bibinfo  {journal} {Rept. Prog. Phys.}\ }\textbf
  {\bibinfo {volume} {79}},\ \bibinfo {pages} {106901} (\bibinfo {year}
  {2016})},\ \Eprint {https://arxiv.org/abs/1511.07586} {arXiv:1511.07586
  [gr-qc]} \BibitemShut {NoStop}%
\bibitem [{\citenamefont {Krssak}\ \emph {et~al.}(2019)\citenamefont {Krssak},
  \citenamefont {van~den Hoogen}, \citenamefont {Pereira}, \citenamefont
  {B{\"o}hmer},\ and\ \citenamefont {Coley}}]{Krssak:2018ywd}%
  \BibitemOpen
  \bibfield  {author} {\bibinfo {author} {\bibfnamefont {M.}~\bibnamefont
  {Krssak}}, \bibinfo {author} {\bibfnamefont {R.~J.}\ \bibnamefont {van~den
  Hoogen}}, \bibinfo {author} {\bibfnamefont {J.~G.}\ \bibnamefont {Pereira}},
  \bibinfo {author} {\bibfnamefont {C.~G.}\ \bibnamefont {B{\"o}hmer}},\ and\
  \bibinfo {author} {\bibfnamefont {A.~A.}\ \bibnamefont {Coley}},\ }\bibfield
  {title} {\bibinfo {title} {{Teleparallel theories of gravity: illuminating a
  fully invariant approach}},\ }\href
  {https://doi.org/10.1088/1361-6382/ab2e1f} {\bibfield  {journal} {\bibinfo
  {journal} {Class. Quant. Grav.}\ }\textbf {\bibinfo {volume} {36}},\ \bibinfo
  {pages} {183001} (\bibinfo {year} {2019})},\ \Eprint
  {https://arxiv.org/abs/1810.12932} {arXiv:1810.12932 [gr-qc]} \BibitemShut
  {NoStop}%
\bibitem [{\citenamefont {Huguet}\ \emph {et~al.}(2021)\citenamefont {Huguet},
  \citenamefont {Le~Delliou}, \citenamefont {Fontanini},\ and\ \citenamefont
  {Lin}}]{Huguet:2020ler}%
  \BibitemOpen
  \bibfield  {author} {\bibinfo {author} {\bibfnamefont {E.}~\bibnamefont
  {Huguet}}, \bibinfo {author} {\bibfnamefont {M.}~\bibnamefont {Le~Delliou}},
  \bibinfo {author} {\bibfnamefont {M.}~\bibnamefont {Fontanini}},\ and\
  \bibinfo {author} {\bibfnamefont {Z.~C.}\ \bibnamefont {Lin}},\ }\bibfield
  {title} {\bibinfo {title} {{Teleparallel gravity as a gauge theory: Coupling
  to matter using the Cartan connection}},\ }\href
  {https://doi.org/10.1103/PhysRevD.103.044061} {\bibfield  {journal} {\bibinfo
   {journal} {Phys. Rev. D}\ }\textbf {\bibinfo {volume} {103}},\ \bibinfo
  {pages} {044061} (\bibinfo {year} {2021})},\ \Eprint
  {https://arxiv.org/abs/2008.13493} {2008.13493 [gr-qc]} \BibitemShut
  {NoStop}%
\bibitem [{\citenamefont {Bahamonde}\ \emph {et~al.}(2023)\citenamefont
  {Bahamonde}, \citenamefont {Dialektopoulos}, \citenamefont
  {Escamilla-Rivera}, \citenamefont {Farrugia}, \citenamefont {Gakis},
  \citenamefont {Hendry}, \citenamefont {Hohmann}, \citenamefont {Levi~Said},
  \citenamefont {Mifsud},\ and\ \citenamefont
  {Di~Valentino}}]{Bahamonde:2021gfp}%
  \BibitemOpen
  \bibfield  {author} {\bibinfo {author} {\bibfnamefont {S.}~\bibnamefont
  {Bahamonde}}, \bibinfo {author} {\bibfnamefont {K.~F.}\ \bibnamefont
  {Dialektopoulos}}, \bibinfo {author} {\bibfnamefont {C.}~\bibnamefont
  {Escamilla-Rivera}}, \bibinfo {author} {\bibfnamefont {G.}~\bibnamefont
  {Farrugia}}, \bibinfo {author} {\bibfnamefont {V.}~\bibnamefont {Gakis}},
  \bibinfo {author} {\bibfnamefont {M.}~\bibnamefont {Hendry}}, \bibinfo
  {author} {\bibfnamefont {M.}~\bibnamefont {Hohmann}}, \bibinfo {author}
  {\bibfnamefont {J.}~\bibnamefont {Levi~Said}}, \bibinfo {author}
  {\bibfnamefont {J.}~\bibnamefont {Mifsud}},\ and\ \bibinfo {author}
  {\bibfnamefont {E.}~\bibnamefont {Di~Valentino}},\ }\bibfield  {title}
  {\bibinfo {title} {{Teleparallel gravity: from theory to cosmology}},\ }\href
  {https://doi.org/10.1088/1361-6633/ac9cef} {\bibfield  {journal} {\bibinfo
  {journal} {Rept. Prog. Phys.}\ }\textbf {\bibinfo {volume} {86}},\ \bibinfo
  {pages} {026901} (\bibinfo {year} {2023})},\ \Eprint
  {https://arxiv.org/abs/2106.13793} {arXiv:2106.13793 [gr-qc]} \BibitemShut
  {NoStop}%
\bibitem [{\citenamefont {Tan}\ \emph {et~al.}(2021)\citenamefont {Tan},
  \citenamefont {Guo}, \citenamefont {Zhang},\ and\ \citenamefont
  {Liu}}]{Tan:2020sys}%
  \BibitemOpen
  \bibfield  {author} {\bibinfo {author} {\bibfnamefont {Q.}~\bibnamefont
  {Tan}}, \bibinfo {author} {\bibfnamefont {W.-D.}\ \bibnamefont {Guo}},
  \bibinfo {author} {\bibfnamefont {Y.-P.}\ \bibnamefont {Zhang}},\ and\
  \bibinfo {author} {\bibfnamefont {Y.-X.}\ \bibnamefont {Liu}},\ }\bibfield
  {title} {\bibinfo {title} {{Gravitational resonances on $f(T)$-branes}},\
  }\href {https://doi.org/10.1140/epjc/s10052-021-09162-0} {\bibfield
  {journal} {\bibinfo  {journal} {Eur. Phys. J. C}\ }\textbf {\bibinfo {volume}
  {81}},\ \bibinfo {pages} {373} (\bibinfo {year} {2021})},\ \Eprint
  {https://arxiv.org/abs/2008.08440} {arXiv:2008.08440 [gr-qc]} \BibitemShut
  {NoStop}%
\bibitem [{\citenamefont {Melfo}\ \emph {et~al.}(2003)\citenamefont {Melfo},
  \citenamefont {Pantoja},\ and\ \citenamefont {Skirzewski}}]{Melfo:2002}%
  \BibitemOpen
  \bibfield  {author} {\bibinfo {author} {\bibfnamefont {A.}~\bibnamefont
  {Melfo}}, \bibinfo {author} {\bibfnamefont {N.}~\bibnamefont {Pantoja}},\
  and\ \bibinfo {author} {\bibfnamefont {A.}~\bibnamefont {Skirzewski}},\
  }\bibfield  {title} {\bibinfo {title} {{Thick domain wall space-times with
  and without reflection symmetry}},\ }\href
  {https://doi.org/10.1103/PhysRevD.67.105003} {\bibfield  {journal} {\bibinfo
  {journal} {Phys. Rev. D}\ }\textbf {\bibinfo {volume} {67}},\ \bibinfo
  {pages} {105003} (\bibinfo {year} {2003})},\ \Eprint
  {https://arxiv.org/abs/gr-qc/0211081} {arXiv:gr-qc/0211081} \BibitemShut
  {NoStop}%
\bibitem [{\citenamefont {Guo}\ \emph {et~al.}(2016)\citenamefont {Guo},
  \citenamefont {Fu}, \citenamefont {Zhang},\ and\ \citenamefont
  {Liu}}]{guo2015}%
  \BibitemOpen
  \bibfield  {author} {\bibinfo {author} {\bibfnamefont {W.-D.}\ \bibnamefont
  {Guo}}, \bibinfo {author} {\bibfnamefont {Q.-M.}\ \bibnamefont {Fu}},
  \bibinfo {author} {\bibfnamefont {Y.-P.}\ \bibnamefont {Zhang}},\ and\
  \bibinfo {author} {\bibfnamefont {Y.-X.}\ \bibnamefont {Liu}},\ }\bibfield
  {title} {\bibinfo {title} {{Tensor perturbations of $f(T)$-branes}},\ }\href
  {https://doi.org/10.1103/PhysRevD.93.044002} {\bibfield  {journal} {\bibinfo
  {journal} {Phys. Rev. D}\ }\textbf {\bibinfo {volume} {93}},\ \bibinfo
  {pages} {044002} (\bibinfo {year} {2016})},\ \Eprint
  {https://arxiv.org/abs/1511.07143} {arXiv:1511.07143 [hep-th]} \BibitemShut
  {NoStop}%
\bibitem [{\citenamefont {Cooper}\ \emph {et~al.}(1995)\citenamefont {Cooper},
  \citenamefont {Khare},\ and\ \citenamefont {Sukhatme}}]{Cooper:1994eh}%
  \BibitemOpen
  \bibfield  {author} {\bibinfo {author} {\bibfnamefont {F.}~\bibnamefont
  {Cooper}}, \bibinfo {author} {\bibfnamefont {A.}~\bibnamefont {Khare}},\ and\
  \bibinfo {author} {\bibfnamefont {U.}~\bibnamefont {Sukhatme}},\ }\bibfield
  {title} {\bibinfo {title} {{Supersymmetry and quantum mechanics}},\ }\href
  {https://doi.org/10.1016/0370-1573(94)00080-M} {\bibfield  {journal}
  {\bibinfo  {journal} {Phys. Rept.}\ }\textbf {\bibinfo {volume} {251}},\
  \bibinfo {pages} {267} (\bibinfo {year} {1995})},\ \Eprint
  {https://arxiv.org/abs/hep-th/9405029} {arXiv:hep-th/9405029} \BibitemShut
  {NoStop}%
\bibitem [{\citenamefont {Ge}\ \emph {et~al.}(2019)\citenamefont {Ge},
  \citenamefont {Jiang}, \citenamefont {Wang}, \citenamefont {Zhang},\ and\
  \citenamefont {Zhong}}]{Ge:2018vjq}%
  \BibitemOpen
  \bibfield  {author} {\bibinfo {author} {\bibfnamefont {B.}~\bibnamefont
  {Ge}}, \bibinfo {author} {\bibfnamefont {J.}~\bibnamefont {Jiang}}, \bibinfo
  {author} {\bibfnamefont {B.}~\bibnamefont {Wang}}, \bibinfo {author}
  {\bibfnamefont {H.}~\bibnamefont {Zhang}},\ and\ \bibinfo {author}
  {\bibfnamefont {Z.}~\bibnamefont {Zhong}},\ }\bibfield  {title} {\bibinfo
  {title} {{Strong cosmic censorship for the massless Dirac field in the
  Reissner-Nordstrom-de Sitter spacetime}},\ }\href
  {https://doi.org/10.1007/JHEP01(2019)123} {\bibfield  {journal} {\bibinfo
  {journal} {JHEP}\ }\textbf {\bibinfo {volume} {01}},\ \bibinfo {pages}
  {123}},\ \Eprint {https://arxiv.org/abs/1810.12128} {arXiv:1810.12128
  [gr-qc]} \BibitemShut {NoStop}%
\bibitem [{\citenamefont {Ciftci}\ \emph {et~al.}(2005)\citenamefont {Ciftci},
  \citenamefont {Hall},\ and\ \citenamefont {Saad}}]{AIM_2005}%
  \BibitemOpen
  \bibfield  {author} {\bibinfo {author} {\bibfnamefont {H.}~\bibnamefont
  {Ciftci}}, \bibinfo {author} {\bibfnamefont {R.~L.}\ \bibnamefont {Hall}},\
  and\ \bibinfo {author} {\bibfnamefont {N.}~\bibnamefont {Saad}},\ }\bibfield
  {title} {\bibinfo {title} {Construction of exact solutions to eigenvalue
  problems by the asymptotic iteration method},\ }\href
  {https://doi.org/10.1088/0305-4470/38/5/015} {\bibfield  {journal} {\bibinfo
  {journal} {Journal of Physics A: Mathematical and General}\ }\textbf
  {\bibinfo {volume} {38}},\ \bibinfo {pages} {1147–1155} (\bibinfo {year}
  {2005})}\BibitemShut {NoStop}%
\bibitem [{\citenamefont {Fortuna}\ and\ \citenamefont
  {Vega}(2023)}]{BSM_2023}%
  \BibitemOpen
  \bibfield  {author} {\bibinfo {author} {\bibfnamefont {S.}~\bibnamefont
  {Fortuna}}\ and\ \bibinfo {author} {\bibfnamefont {I.}~\bibnamefont {Vega}},\
  }\bibfield  {title} {\bibinfo {title} {{Bernstein spectral method for
  quasinormal modes and other eigenvalue problems}},\ }\href
  {https://doi.org/10.1140/epjc/s10052-023-12350-9} {\bibfield  {journal}
  {\bibinfo  {journal} {Eur. Phys. J. C}\ }\textbf {\bibinfo {volume} {83}},\
  \bibinfo {pages} {1170} (\bibinfo {year} {2023})},\ \Eprint
  {https://arxiv.org/abs/2003.06232} {arXiv:2003.06232 [gr-qc]} \BibitemShut
  {NoStop}%
\bibitem [{\citenamefont {Pani}(2013)}]{DIM_2013}%
  \BibitemOpen
  \bibfield  {author} {\bibinfo {author} {\bibfnamefont {P.}~\bibnamefont
  {Pani}},\ }\bibfield  {title} {\bibinfo {title} {{Advanced Methods in
  Black-Hole Perturbation Theory}},\ }\href
  {https://doi.org/10.1142/S0217751X13400186} {\bibfield  {journal} {\bibinfo
  {journal} {Int. J. Mod. Phys. A}\ }\textbf {\bibinfo {volume} {28}},\
  \bibinfo {pages} {1340018} (\bibinfo {year} {2013})},\ \Eprint
  {https://arxiv.org/abs/1305.6759} {arXiv:1305.6759 [gr-qc]} \BibitemShut
  {NoStop}%
\bibitem [{\citenamefont {Cho}\ \emph {et~al.}(2012)\citenamefont {Cho},
  \citenamefont {Cornell}, \citenamefont {Doukas}, \citenamefont {Huang},\ and\
  \citenamefont {Naylor}}]{AIM_2011}%
  \BibitemOpen
  \bibfield  {author} {\bibinfo {author} {\bibfnamefont {H.~T.}\ \bibnamefont
  {Cho}}, \bibinfo {author} {\bibfnamefont {A.~S.}\ \bibnamefont {Cornell}},
  \bibinfo {author} {\bibfnamefont {J.}~\bibnamefont {Doukas}}, \bibinfo
  {author} {\bibfnamefont {T.~R.}\ \bibnamefont {Huang}},\ and\ \bibinfo
  {author} {\bibfnamefont {W.}~\bibnamefont {Naylor}},\ }\bibfield  {title}
  {\bibinfo {title} {{A New Approach to Black Hole Quasinormal Modes: A Review
  of the Asymptotic Iteration Method}},\ }\href
  {https://doi.org/10.1155/2012/281705} {\bibfield  {journal} {\bibinfo
  {journal} {Adv. Math. Phys.}\ }\textbf {\bibinfo {volume} {2012}},\ \bibinfo
  {pages} {281705} (\bibinfo {year} {2012})},\ \Eprint
  {https://arxiv.org/abs/1111.5024} {arXiv:1111.5024 [gr-qc]} \BibitemShut
  {NoStop}%
\end{thebibliography}%

\end{document}